# Quantization and Biphoton Statistics of k-Gap Solitons in Nonlinear Photonic Time Crystals

Liang Zhang, Chenhao Pan, Yiming Pan*

State Key Laboratory of Quantum Functional Materials, School of Physical Science and Technology and Center for Transformative Science, ShanghaiTech University, Shanghai 200031, China

**Abstract**

Nonlinear photonic time crystals (PTCs) can support solitons inside momentum k-gaps, where the amplification of k-gap modes is saturated by Kerr nonlinearity, forming spatially homogeneous but temporally localized excitations. Yet their quantum nature remains unclear. Here we quantize nonlinear k-gap dynamics of PTCs and show that k-gap solitons are represented by biphoton Fock-ladder states. K-gap amplification drives two-mode squeezing of the biphoton, while Kerr nonlinearity generates an anharmonic potential along the biphoton-Fock ladder that balances this squeezing process, creating a finite biphoton-number turning point and giving rise to quantum collapse–revival dynamics and nonclassical phase-space interference. We further analyze how photon loss and dephasing reshape the biphoton statistics of quantized k-gap solitons. Our results establish a biphoton Fock-space description of k-gap soliton quantization and provide a framework for studying quantum nonlinear excitations and entangled light generation in photonic time crystals.

Solitons exhibit a wave–particle dual character: they are wave-like in origin, yet strikingly particle-like in their persistence during propagation and interactions[1]. They arise when dispersion or diffraction is balanced by nonlinearity, enabling waveforms that persist over long distances[2,3]. In periodic media, this balancing principle underlies gap solitons, including Bragg solitons, where nonlinearity sustains self-localized states inside linear bandgaps[4–6]. Nonlinear photonic time crystals (PTCs) extend this mechanism to media periodic in time rather than space[7]. Their temporal modulation on dielectric constants opens momentum bandgaps, or k-gaps, and enables photon-pair generation and exponential amplification of k-gap modes[8]. In this setting, k-gap solitons were recently identified as nonlinear excitations that reside in a momentum gap, remain plane-wave-like in space, and localize only in time through the competition between k-gap amplification and nonlinear suppression[9]. However, while the classical formation mechanism of k-gap solitons has been identified, their quantum counterpart remains unexplored.

Quantization of optical solitons is generally a multimode quantum-field problem, as exemplified by optical-fiber solitons. In the treatments of Drummond and co-workers[10,11], solitons are described from a fully quantized nonlinear field, whereas Lai and Haus approached this problem from complementary perspectives, using the time-dependent Hartree approximation and exact Bethe-ansatz soliton states[12,13]. These studies established that a quantum soliton is not a single bosonic mode, but a localized many-component field state. Upon quantization, different field components acquire distinct phases and fluctuations, replacing the classical picture of shape-preserving propagation with phase diffusion, wave-packet spreading, squeezing, and collision-induced correlations.

The present problem is distinct because the classical PTC dynamics is already preselected near the k-gap center. Modes at the gap center have the largest growth rate, so k-gap amplification acts as a mode filter, rapidly selecting a dominant $(k, -k)$ pair while suppressing more weakly amplified components or on-band Floquet modes. After amplification, the nonlinear evolution therefore develops not across a broad continuum, but within an effectively monochromatic two-mode manifold. Upon quantization of the linear PTC dynamics[8], the modulation creates or annihilates only opposite-momentum photon pairs, conserving the photon-number difference between the two modes. Seeding from quantum vacuum, the dynamics is restricted to the biphoton Fock-ladder $|0,0\rangle, |1,1\rangle, |2,2\rangle, \ldots$, in a nonlinear PTC, the quantized Kerr response preserves this paired-ladder structure while generating a number-dependent anharmonic potential along it. The k-gap soliton is therefore quantized as a reduced biphoton Fock-ladder problem, rather than as a generic multimode continuum-field soliton.

Here we develop a quantum description of k-gap solitons starting from the reduced coupled-mode equations of a nonlinear PTC. By quantizing the dominant opposite-momentum mode pair selected by the k-gap, we derive a two-mode Hamiltonian and express the vacuum-initiated dynamics in the paired Fock basis. This formulation identifies how the classical balance between k-gap amplification and Kerr saturation is encoded in the quantum problem. We then analyze the resulting biphoton-number dynamics. We further characterize the quantized $k$-gap soliton using a biphoton phase-space representation and Hanbury Brown–Twiss correlations, which provide experimentally accessible signatures of biphoton statistics. Our results establish a biphoton Fock-space framework for the quantization of $k$-gap solitons and show how soliton physics in nonlinear time-modulated media can engage with quantum light generation.

**k-gap solitons.** We consider a nonlinear photonic time crystal described as a spatially homogeneous Kerr medium with temporally periodic permittivity, as shown in Fig. 1a. Applying the slowly varying envelope approximation to Maxwell's equations yields reduced coupled-mode equations for the two resonantly coupled opposite-momentum envelopes, $A_k$ and $A_{-k}$ (Supplementary Note 1):

$$
\begin{aligned}
i\frac{dA_k}{dt} &= [\delta(k)+\gamma(|A_k|^2+2|A_{-k}|^2)]A_k + \kappa A_{-k}^* \\
i\frac{dA_{-k}}{dt} &= [\delta(k)+\gamma(|A_{-k}|^2+2|A_k|^2)]A_{-k} + \kappa A_k^*
\end{aligned}
\qquad (1)
$$

Here, $\delta(k)=c(|k|-k_0)$ is the detuning from the center of the frequency resonance, with $k_0=\Omega/2c$. Temporal modulation couples the two modes with strength $\kappa=\Delta\epsilon*\Omega/8$, where $\Delta\epsilon \ll 1$ is the dimensionless modulation depth and $\Omega=2\pi/\mathrm{T}$ is the modulation frequency. The nonlinear response is taken to be instantaneous Kerr, with $\gamma=-\beta\Omega/4$ and $\beta=3\chi_3/\epsilon_0\epsilon_r^3$. Although we focus on Kerr nonlinearity, the formalism can be extended to other local nonlinear responses. Equation (1) captures the competition between temporal mode coupling and nonlinear phase modulation, providing the classical starting point for k-gap-soliton dynamics.

To obtain the linear k-gap spectrum, we first set $\gamma=0$. The linearized equations lead to a Bogoliubov-de Gennes eigenvalue problem, with eigenfrequencies relative to the gap center given by $\omega_\pm(k)=\omega_0\pm\sqrt{\delta(k)^2-\kappa^2}$. As shown in Fig. 1b, this dispersion contains two symmetric momentum gaps centered at $k=\pm k_0$. Outside the gap, the eigenfrequencies are real and the on-band modes oscillate. Inside the gaps, $|\delta(k)|<\kappa$, the frequencies become complex, $\omega(k)=\omega_0\pm i\Gamma(k)$, with $\Gamma(k)=\sqrt{\kappa^2-\delta(k)^2}$. The corresponding modes therefore grow or decay exponentially in time. The growth rate is maximal at the gap center, $\delta(k)=0$, where $\Gamma_{max}=\kappa$. Away from the gap center, the relative growth of an off-center mode compared with the resonant pair is reduced by the factor $\exp[-(\Gamma_{max}-\Gamma(k))t]$. Near $k_0$, this gain difference scales as $\Gamma_{max}-\Gamma(k)\simeq\delta(k)^2/(2\kappa)$. Thus, k-gap amplification acts as a mode filter, rapidly selecting the phase-matched $(k_0,-k_0)$ pair while leaving off-center modes secondary, and the validity of this dominant-pair description is quantified in Supplementary Note 2. Thus, in the nonlinear analysis below, we focus on these selected gap-center pairs and set $\delta = 0$.

The two-mode equations (1) conserve the intensity imbalance $D = |A_k|^2 - |A_{-k}|^2$. For equal initial intensities, the dynamics remains on the symmetric manifold $|A_k| = |A_{-k}|$. Writing $A_k = \sqrt{I}e^{i\phi_k}, A_{-k} = \sqrt{I}e^{i\phi_{-k}}$, and defining the pair phase $\theta = \phi_k + \phi_{-k}$, Eq. (1) reduces to $\frac{dI}{dt} = -2\kappa I \sin\theta, \frac{d\theta}{dt} = -2\kappa\cos\theta - 6\gamma I$. This reduced system conserves $H(I,\theta) = 2\kappa I\cos\theta + 3\gamma I^2 = E$. Eliminating $\theta$ gives $\left(\frac{dI}{dt}\right)^2 = 4\kappa^2 I^2 - (E - 3\gamma I^2)^2$. For $E = 0$, the intensity takes the localized form

$$I(t) = \frac{2\kappa}{3\gamma\cosh[2\kappa(t - t_0)]}, \tag{2}$$

with peak intensity $I_{max} = \frac{2\kappa}{3\gamma}$. Equation (2) shows that the classical $k$-gap soliton results from the balance between $k$-gap amplification ($\kappa$) and Kerr nonlinear phase modulation ($\gamma$). Linear $k$-gap amplification drives exponential growth, whereas the Kerr-induced phase shift suppresses this growth and saturates it into a transient, temporally localized peak. The same soliton solution can also be obtained by reducing the coupled-mode equations to an effective nonlinear Schrödinger equation for the slowly varying envelope[9].

**Quantization of k-gap solitons.** The classical analysis above shows that the dominant nonlinear dynamics is carried by the phase-matched $(k_0, -k_0)$ pair selected by k-gap amplification. We therefore quantize this selected two-mode manifold by promoting the classical amplitudes to bosonic operators[14,15], $A_{\pm k} \to A_0\hat{a}_{\pm k}$ and $A^*_{\pm k} \to A^*_0\hat{a}^\dagger_{\pm k}$, with $[\hat{a}_j, \hat{a}^\dagger_l] = \delta_{jl}$ and $\hat{n}_{\pm k} = \hat{a}^\dagger_{\pm k}\hat{a}_{\pm k}$, and $|A_0| = |A^*_0|$ being the quantized mode strength. The effective Hamiltonian is chosen so that the Heisenberg equations of motion, $i\hbar\frac{d\hat{a}_j}{dt} = [\hat{a}_j, \hat{H}_{\text{eff}}]$, recover the reduced coupled-mode equations in the classical mean-field limit. At the $k$-gap center, where the detuning vanishes, this gives

$$\hat{H}_{\text{eff}} = \hbar\tilde{\kappa}\left(\hat{a}^\dagger_k\,\hat{a}^\dagger_{-k} + \hat{a}_k\,\hat{a}_{-k}\right) + \frac{\hbar U}{2}\left[\hat{n}^2_k + \hat{n}^2_{-k} + 4\hat{n}_k\,\hat{n}_{-k} - (\hat{n}_k + \hat{n}_{-k})\right], \tag{3}$$

The derivation of Eq. (3), including the Heisenberg equations and normal ordering of the self- and cross-Kerr terms, is given in Supplementary Note 3. Here, $U$ is the quantized Kerr coefficient corresponding to the classical nonlinear coefficient $|A_0|^2\gamma$. The first term is the quantum form of the temporal-modulation coupling, with coefficient as $\tilde{\kappa} = \kappa$. Unlike a beam-splitter interaction, $\hat{a}_k^\dagger \hat{a}_{-k} + h.c.$, it creates or annihilates photons simultaneously in the two opposite-momentum modes through $\hat{a}_k^\dagger \hat{a}_{-k}^\dagger + \hat{a}_k \hat{a}_{-k}$. The second term is the normally ordered Kerr interaction in number-operator form: the self-Kerr contribution gives $\hat{n}_{\pm k}^2 - \hat{n}_{\pm k}$, while the cross-Kerr term gives $4\hat{n}_k \hat{n}_{-k}$.

Because the pair term in Eq. (3) creates or annihilates photons simultaneously in the two opposite-momentum modes, the photon-number difference $\hat{Q} = \hat{n}_k - \hat{n}_{-k}$ is conserved, equivalently $[\hat{Q}, \hat{H}_{\mathrm{eff}}] = 0$. The Hilbert space therefore decomposes into independent Q-resolved biphoton-Fock ladders. We use "biphoton" to denote a photon pair occupying the $(k, -k)$ modes[16]. Starting from vacuum, the system remains in the $Q = 0$ sector and is restricted to the biphoton Fock ladder $|0,0\rangle$, $|1,1\rangle$, $|2,2\rangle, \ldots, |n,n\rangle$, ..., as illustrated in Fig. 2a. The state can therefore be written as $|\psi(t)\rangle = \sum_{n=0}^{\infty} c_n(t)|n,n\rangle$, where $n$ denotes the biphoton number. Acting with Eq. (3) on this basis gives

$$\hat{H}_{\mathrm{eff}}|n,n\rangle = \hbar U(3n^2 - n)|n,n\rangle + \hbar\tilde{\kappa}(n+1)|n+1,n+1\rangle + \hbar\tilde{\kappa}n|n-1,n-1\rangle. \quad (4)$$

Equation (4) identifies the biphoton Fock ladder as a one-dimensional lattice in pair-number space. The diagonal term, $U(3n^2 - n)$, gives the Kerr-induced energy of the state with biphoton number $n$. Thus, the classical Kerr nonlinear phase shift becomes an on-site energy along the biphoton axis, whose quadratic dependence on $n$ forms an anharmonic Fock-space potential. The off-diagonal terms describe hopping between neighboring biphoton states: pair creation connects $|n,n\rangle$ to $|n+1,n+1\rangle$ with amplitude $\tilde{\kappa}(n+1)$, while pair annihilation connects it to $|n-1,n-1\rangle$ with $\tilde{\kappa}n$. The quantized dynamics is therefore mapped onto a biphoton Fock ladder with $n$-dependent hopping and a Kerr-induced anharmonic potential.

Substituting the expansion into the Schrödinger equation $i\hbar\, \partial_t|\psi\rangle = \hat{H}_{\mathrm{eff}}|\psi\rangle$ gives

$$i\frac{dc_n}{dt} = U(3n^2 - n)c_n + \tilde{\kappa}nc_{n-1} + \tilde{\kappa}(n+1)c_{n+1}. \quad (5)$$

with $c_{-1} = 0$. The derivation of the biphoton Fock-ladder equation is provided in Supplementary Note 3. Equation (5) describes evolution along the discrete biphoton coordinate $n$, as schematically shown in Fig. 2a. To highlight pair generation (as calibration), we first set $U = 0$. For the vacuum initial state, $c_0(0) = 1$ and $c_{n>0}(0) = 0$, the temporal coupling progressively populates higher biphoton states. This ladder dynamic then has the exact solution $c_n(t) = \mathrm{sech}(\tilde{\kappa}t)\,[-i\tanh(\tilde{\kappa}t)]^n$, and hence the biphoton statistics $P_{\mathrm{n}}(t) = |c_{\mathrm{n}}(t)|^2 = \mathrm{sech}^2(\tilde{\kappa}t)\tanh^{2\mathrm{n}}(\tilde{\kappa}t)$. This geometric distribution has mean biphoton number $\langle n\rangle = \sum_n nP_n(t) = \sinh^2(\tilde{\kappa}t)$. Thus, in the absence of Kerr nonlinearity, vacuum evolves into biphoton statistics whose mean population follows the same hyperbolic growth as linear k-gap amplification, providing the two-mode squeezed reference dynamics[17,18].

We next consider Eq. (5) at finite Kerr nonlinearity. At early times, the biphoton distribution remains concentrated near small $n$, where the Kerr-induced diagonal term $U(3n^2 - n)$ is weak. The dynamics is then dominated by pair hopping along the biphoton ladder. Because the hopping amplitudes scale as $\tilde{\kappa}(n+1)$ for $|n,n\rangle \to |n+1,n+1\rangle$ and $\tilde{\kappa}n$ for $|n,n\rangle \to |n-1,n-1\rangle$, vacuum seeding drives the distribution away from $|0,0\rangle$ toward larger $n$, producing the initial expansion that closely follows the linear two-mode-squeezing reference [Fig. 2b]. In this sense, the early biphoton-Fock expansion is the quantum counterpart of the exponential k-gap amplification that drives the initial growth of classical k-gap solitons.

At larger biphoton numbers, the Kerr term becomes important. The diagonal energy $U(3n^2 - n)$ forms an anharmonic, approximately quadratic potential along the biphoton-number axis. Since this potential grows as $n^2$, whereas the hopping scale grows only linearly with $n$, the outward expansion is arrested when the Kerr-induced detuning suppresses further hopping toward higher pair-number states. We identify this region as the Fock-space turning point $n_{tp}$, the largest biphoton number reached before nonlinearity saturation. To estimate $n_{tp}$, we use a semiclassical form of Eq. (5). In the large-$n$ limit, $n$ is treated as a continuous coordinate and the phase difference between neighboring Fock states is denoted by $p$, such that $c_{n\pm1} \sim c_n e^{\pm ip}$. Equation (5) then gives an effective Fock-space Hamiltonian

$$H(n,p) = 2\tilde{\kappa}n\cos p + U(3n^2 - n). \tag{6}$$

For the trajectory connected to the vacuum state, $H = 0$. At the outer boundary of the allowed motion, the distribution reaches its largest accessible pair number, corresponding to $p = \pi$. This gives $2\tilde{\kappa}n_{tp} = U(3n_{tp}^2 - n_{tp})$, and in the large-$n$ limit, $n_{tp} \simeq \frac{2\tilde{\kappa}}{3U}$. For the parameters in Fig. 2, $\tilde{\kappa} = 1$ and $U = 0.02$, this gives $n_{tp} \simeq 33$, consistent with the direct simulation result of the turning region in Fig.2c. This estimates also links the Fock-space turning point to the classical soliton peak. Using $I \sim A_0^2 n$, $U = \gamma A_0^2$, and $\tilde{\kappa} = \kappa$, we obtain $A_0^2 n_{\mathrm{tp}} \simeq \frac{2\kappa}{3\gamma} = I_{\mathrm{max}}$. Thus, $n_{tp}$ plays in Fock space the role that $I_{max}$ plays in the classical theory: $I_{max}$ is the maximum intensity of k-gap soliton, whereas $n_{tp}$ is the largest biphoton state reached by its quantized counterpart.

After reaching $n_{\mathrm{tp}}$, the biphoton distribution can no longer keep expanding toward higher $n$. The Kerr-induced potential redirects it toward lower pair-number states, producing the collapse stage shown in Fig. 2d. As the distribution returns to lower $n$, the potential wakens and hopping becomes effective again, pushing the distribution toward larger $n$, giving rise to the revival stage in Fig. 2e. The mean pair number $\langle n \rangle$ in Fig. 2f captures the full squeezing–turning–collapse–revival cycle, distinguishing Kerr-limited k-gap dynamics from unbounded linear two-mode squeezing.

The biphoton-number distribution $P(n,t) = |c_n(t)|^2$ tracks the population flow along the $Q = 0$ Fock ladder. As shown in Fig. 3a, the distribution expands from vacuum to larger biphoton numbers, reaches a turning region, and then collapses back toward lower $n$. At later times, different Fock states accumulate different dynamical phases because the Kerr energy $U(3n^2 - n)$ depends on $n$. This relative phase spreading, which we refer to as biphoton phase diffusion, appears in Fig. 3a as a blurred late-time population pattern and show that $P(n,t)$ alone does not fully characterize the quantum dynamics. The phase diffusion leads to the squeezing–turning–collapse–revival cycle, similar to the collapse–revival dynamics of quantized Rabi oscillation[19,20]. The phase diffusion determines the off-diagonal density-matrix elements, $\rho_{nm} = c_n c_m^* = |c_n||c_m|e^{i(\phi_n - \phi_m)}$, which encode phase coherence and interference between biphoton states. To visualize this coherence, we introduce a Wigner representation for the one-dimensional biphoton Fock-ladder[21–23]. Denoting the paired state $|n,n\rangle$ as $|n\rangle$, we define

$$W(X,P;t) = \sum_{n,m} \rho_{nm}(t) W_{nm}(X,P)\,, \tag{7}$$

where $W_{nm}(X,P)$ is the phase-space representation of $|n\rangle\langle m|$. The diagonal terms $W_{nn}$ reflect biphoton-state occupations, whereas the off-diagonal terms $W_{nm}$ encode phase coherence and interference. The coordinates $(X,P)$ are therefore quadratures of the biphoton ladder, not the conventional quadrature of the optical field. The algebraic construction is given in Supplementary Note 6.

In the ideal case, the Wigner representation resolves the phase-space evolution of the three stages marked in Fig. 3a. The marginal distribution $\mathcal{P}(X) = \int W(X,P)\,dP$ gives the corresponding $X$-quadrature profile. At $t_1$, during the squeezing stage, the Wigner function is elongated and tilted, while its weight remains close to the origin; $\mathcal{P}(X)$ is correspondingly smooth and single peaked. At $t_2$, near the turning region, the distribution is displaced toward negative-$X$ and becomes more extended. This short-lived stage coincides with the maximum mean pair number $\langle n \rangle$ in Fig. 2f, and represents the Fock-space counterpart of the classical $k$-gap soliton peak $I_{\max}$. Alternating positive and negative Wigner regions reveal nonclassical interference between biphoton Fock components, while the broadened, structured $\mathcal{P}(X)$ carries this interference into the diagnostic quadrature distribution. At $t_3$, after the population collapses toward lower $n$, the Wigner distribution rotates back toward the low-amplitude region of the effective quadrature phase space while retaining residual fine structures. The recovered narrow peak in $\mathcal{P}(X)$ shows that collapse reflects coherent redistribution of biphoton state, rather than simple population decay.

To assess the effect of dissipation, we include effective photon loss and dephasing in the biphoton Fock ladder. Unlike independent single-photon loss in the $+k$ and $-k$ modes, this effective model acts on the correlated states $|n,n\rangle$, describing dissipation within the paired excitation manifold. As shown in Fig. 3e–g, the dissipative state at $t_1$ remains close to the ideal squeezed profile: the Wigner distribution is centered near the origin and keeps its elongated shape, while $\mathcal{P}(X)$ remains smooth and single-peaked. At $t_2$, the distribution becomes more diffuse and the interference fringes are strongly suppressed; correspondingly, $\mathcal{P}(X)$ loses most of its oscillatory structure. At $t_3$, instead of returning to the original site of the quadrature phase space, the Wigner distribution settles around the center of a dissipatively modified trajectory. The associated $\mathcal{P}(X)$ has a reduced peak height and a clear leftward shift. Photon loss and dephasing therefore reshape the phase-space motion by suppressing fine interference fringes and preventing the near-periodic rotational return of the ideal evolution. Details of the Wigner quadrature representation and phase-space snapshots are provided in Supplementary Note 4.

To connect the Fock-space dynamics to an experimentally accessible observable, we evaluate Hanbury Brown–Twiss (HBT) correlations of the generated biphotons[24,25]. As sketched in Fig. 4, the output from the nonlinear PTC is sent to a 50:50 beam splitter and detected by two photon counters, allowing coincidence statistics to be measured. We focus on the zero-delay correlation $g^{(2)}(t;\tau=0)$, where $t$ is the modulation time. As shown in Fig. 4, $g^{(2)}$ exhibits pronounced time-dependent oscillations that reflect the cycling of squeezing, turning, collapse, and revival. The cross-correlation $g^{(2)}_{+k,-k}(t;0)$ probes the $\pm k$ biphoton emission, while the auto-correlation $g^{(2)}_{+k,+k}(t;0)$ characterizes photon bunching. These photon correlations provide a photon-counting readout of quantized $k$-gap soliton dynamics.

**Further discussion**

The analysis above focuses on vacuum-initiated dynamics. A natural extension is coherent-state seeding, as discussed in Supplementary Note 5. In this case, the initial state already spans a finite range of Fock states, and the evolution describes a seeded quantum field rather than spontaneous biphoton generation from vacuum. These two initial conditions expose complementary limits: vacuum seeding emphasizes Fock-space discreteness, biphoton correlations, and collapse–revival dynamics, whereas coherent seeding connects more directly to the semiclassical mean-field dynamics of classical k-gap solitons.

Soliton quantization reveals how nonlinear self-organization is modified by quantum fluctuations and photon correlations. In conventional optical-soliton theory, the field is usually expanded around a classical soliton background, $\psi = \psi_{\mathrm{soliton}} + \delta\psi$, and the fluctuation field $\delta\psi$ is quantized11,12. This approach captures quantum noise, squeezing, and phase diffusion on top of a classical soliton. The k-gap soliton studied here requires a different viewpoint. Its classical dynamics originates from a Floquet–Bogoliubov pair process with growing and decaying k-gap branches, rather than from a purely conservative soliton background. Since k-gap amplification selects a dominant opposite-momentum pair, we quantize this preselected pair directly, which gives a monochromatic biphoton Fock-ladder description. In this picture, the classical soliton is fully quantized and its quantum nature exhibits collapse–revival behavior and nonclassical biphoton interference. Extending this reduced two-mode framework to multimode solitons, such as event solitons in nonlinear spacetime crystals[26] and broader Floquet continua, remains an open direction.

Experimentally, k-gap amplification and related temporal-topological effects have recently been explored in time-varying metasurfaces[27], microwave transmission-line systems[28], and acoustic or phononic time-crystal platforms[29]. These studies have mainly addressed classical amplification and k-gap physics. At optical frequencies, optically modulated ITO films have demonstrated coherent absorption and amplification, suggesting a promising route toward realistic photonic-time-crystal platforms[30]. Such developments may enable future detection of quantized k-gap soliton features, including biphoton statistics and quantum coherence.

**Conclusion**

We have developed a quantum description of k-gap solitons in nonlinear photonic time crystals by exploiting the mode selection induced by k-gap amplification. The dominant $(k, -k)$ mode pair defines a reduced quantum manifold in which soliton dynamics is represented on a biphoton Fock ladder. In the linear regime, this formulation recovers vacuum-seeded two-mode squeezing. With Kerr nonlinearity, the classical saturation of k-gap amplification is converted into a quadratic potential along the biphoton-Fock ladder, which confines the squeezing and produces collapse–revival behavior and nonclassical phase-space interference. The resulting turning point $n_{tp}$ serves as the nonclassical Fock-space counterpart of the k-gap soliton peak $I_{\max}$, linking the mean-field soliton picture to its quantized biphoton dynamics. We further showed that photon loss and dephasing reshape the effective biphoton phase-space dynamics, while Hanbury Brown–Twiss correlations offer an experimental probe of the biphoton statistics. These results establish a Fock-space picture of quantized k-gap solitons and suggest a route toward connecting nonlinear time-varying media with quantum light generation.

**Acknowledgements**

**Funding**: This work was supported by the NSFC (No. 2023X0201-417-03) and the start-up funding from ShanghaiTech University.

**Authors' contributions:**

**Liang Zhang:** Conceptualization, methodology, numerical simulations, formal analysis, visualization, writing – original draft. **Chenhao Pan:** Discussion and manuscript revision. **Yiming Pan:** Conceptualization, supervision, funding acquisition, writing – review & editing. All authors read and approved the final manuscript.

**Availability of data and materials**

The data that support the findings of this study are available from the corresponding authors upon reasonable request.

**Competing interests**

The authors declare that they have no competing interests.

Correspondence and requests for materials should be addressed to Y.P.

(yiming.pan@shanghaitech.edu.cn)

**Reference**

1. Zabusky, N. J. & Kruskal, M. D. Interaction of" solitons" in a collisionless plasma and the recurrence of initial states. *Phys. Rev. Lett.* **15**, 240 (1965).

2. Hasegawa, A. & Tappert, F. Transmission of stationary nonlinear optical pulses in dispersive dielectric fibers. I. Anomalous dispersion. *Appl. Phys. Lett.* **23**, 142–144 (1973).

3. Mollenauer, L. F., Stolen, R. H. & Gordon, J. P. Experimental observation of picosecond pulse narrowing and solitons in optical fibers. *Phys. Rev. Lett.* **45**, 1095 (1980).

4. Christodoulides, D. N. & Joseph, R. I. Slow bragg solitons in nonlinear periodic structures. *Phys. Rev. Lett.* **62**, 1746 (1989).

5. de Sterke, C. M. & Sipe, J. E. Gap solitons. in *Progress in optics* vol. 33 203–260 (Elsevier, 1994).

6. Eggleton, B. J., Slusher, R. E., De Sterke, C. M., Krug, P. A. & Sipe, J. E. Bragg grating solitons. *Phys. Rev. Lett.* **76**, 1627 (1996).

7. Galiffi, E. *et al.* Photonics of time-varying media. *Advanced Photonics* **4**, 14002 (2022).

8. Lyubarov, M. *et al.* Amplified emission and lasing in photonic time crystals. *Science (1979).* **377**, 425–428 (2022).

9. Pan, Y., Cohen, M.-I. & Segev, M. Superluminal k-gap solitons in nonlinear photonic time crystals. *Phys. Rev. Lett.* **130**, 233801 (2023).

10. Carter, S. J., Drummond, P. D., Reid, M. D. & Shelby, R. M. Squeezing of quantum solitons. *Phys. Rev. Lett.* **58**, 1841 (1987).

11. Drummond, P. D. & Carter, S. J. Quantum-field theory of squeezing in solitons. *Journal of the Optical Society of America B* **4**, 1565–1573 (1987).

12. Lai, Y. & Haus, H. A. Quantum theory of solitons in optical fibers. II. Exact solution. *Phys. Rev. A (Coll. Park).* **40**, 854 (1989).

13. Haus, H. A., Watanabe, K. & Yamamoto, Y. Quantum-nondemolition measurement of optical solitons. *Journal of the Optical Society of America B* **6**, 1138–1148 (1989).

14. Caves, C. M. & Schumaker, B. L. New formalism for two-photon quantum optics. I. Quadrature phases and squeezed states. *Phys. Rev. A (Coll. Park).* **31**, 3068 (1985).

15. Walls, D. F. & Milburn, G. J. *Quantum Optics*. (Springer Science & Business Media, 2008).

16. Shih, Y. *An Introduction to Quantum Optics: Photon and Biphoton Physics*. (CRC press, 2020).

17. Sustaeta-Osuna, J. E., García-Vidal, F. J. & Huidobro, P. A. Quantum theory of photon pair creation in photonic time crystals. *ACS Photonics* **12**, 1873–1880 (2025).

18. Bae, J., Lee, K., Min, B. & Kim, K. W. Quantum electrodynamics of photonic time crystals. *Nat. Commun.* (2025).

19. Jaynes, E. T. & Cummings, F. W. Comparison of quantum and semiclassical radiation theories with application to the beam maser. *Proceedings of the IEEE* **51**, 89–109 (1963).

20. Rempe, G., Walther, H. & Klein, N. Observation of quantum collapse and revival in a one-atom maser. *Phys. Rev. Lett.* **58**, 353 (1987).

21. Wigner, E. On the quantum correction for thermodynamic equilibrium. *Physical review* **40**, 749 (1932).

22. Schleich, W. P. *Quantum Optics in Phase Space*. (John Wiley & Sons, 2015).

23. Leonhardt, U. *Measuring the Quantum State of Light*. vol. 22 (Cambridge university press, 1997).

24. Brown, R. H. & Twiss, R. Q. Correlation between photons in two coherent beams of light.

*Nature* **177**, 27–29 (1956).

25. Glauber, R. J. The quantum theory of optical coherence. *Physical Review* **130**, 2529 (1963).

26. Zhang, L., Fan, Z. & Pan, Y. Event soliton formation in mixed energy-momentum gaps of nonlinear spacetime crystals. *Phys. Rev. Res.* **8**, 023107 (2026).

27. Wang, X. *et al.* Metasurface-based realization of photonic time crystals. *Sci. Adv.* **9**, eadg7541 (2023).

28. Reyes-Ayona, J. R. & Halevi, P. Electromagnetic wave propagation in an externally modulated low-pass transmission line. *IEEE Trans. Microw. Theory Tech.* **64**, 3449–3459 (2016).

29. Tong, S. *et al.* Observation of momentum-band topology in PT-symmetric Floquet lattices. *Nat. Commun.* **16**, 9975 (2025).

30. Galiffi, E. *et al.* Optical coherent perfect absorption and amplification in a time-varying medium. *Nat. Photonics* 1–7 (2026).

**Figures:**

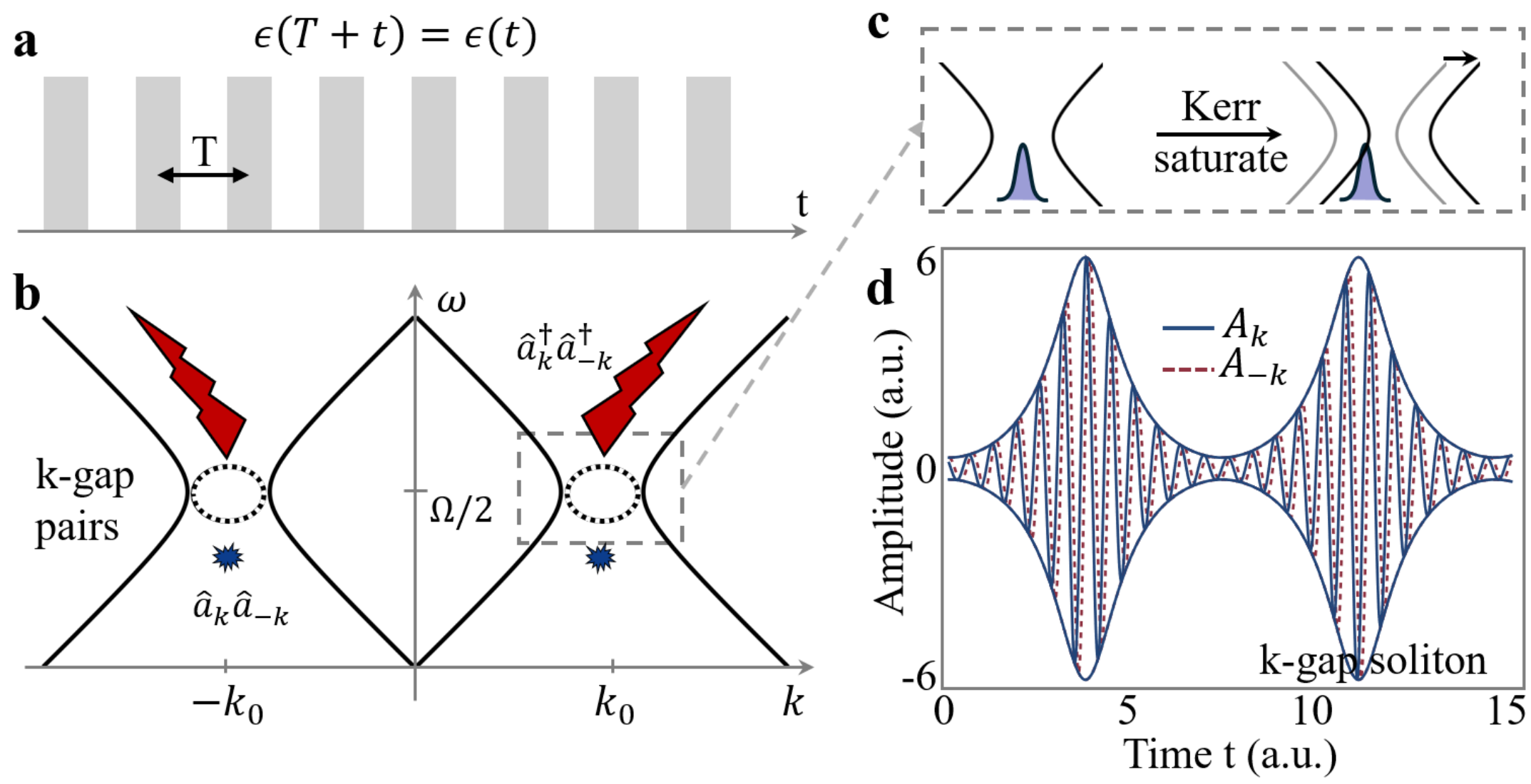

**Figure 1 | Formation of a k-gap soliton in nonlinear photonic time crystals.** (a) Schematic of a photonic time crystal with temporally periodic permittivity, $\epsilon(t+T)=\epsilon(t)$, where $T=2\pi/\Omega$ is the modulation period. (b) Dispersion relation showing momentum gaps opened by temporal modulation near the opposite momenta $\pm k_0$. (c) Kerr nonlinearity saturates k-gap amplification by shifting the effective detuning away from resonance with the injected field at momentum $k_0$. (d) Field evolution of the coupled opposite-momentum modes. The amplitudes $A_k$ and $A_{-k}$ form a temporally localized but spatially uniform $k$-gap soliton.

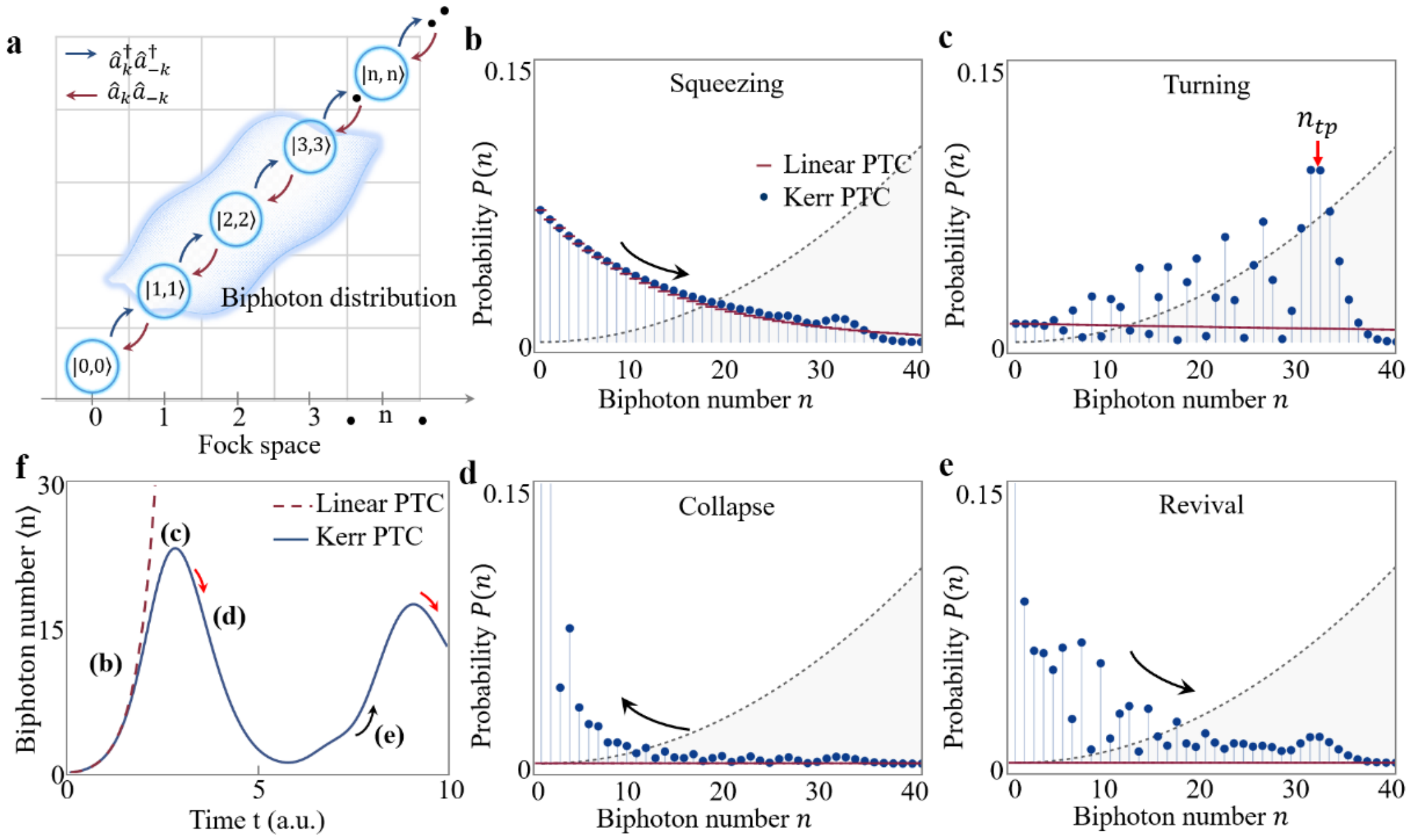

**Figure 2 | Quantum collapse and revival of the k-gap soliton on the biphoton Fock ladder.** (a) The $k$-gap pair process confines the vacuum-initiated dynamics to the biphoton Fock ladder, $|0,0\rangle, |1,1\rangle, \ldots, |n,n\rangle$, ... Pair creation $\hat{a}_k^\dagger \hat{a}_{-k}^\dagger$ and annihilation $\hat{a}_k \hat{a}_{-k}$ drive the biphoton distribution along the paired-photon-number axis. (b–e) Biphoton distributions $P(n)$ at the squeezing, turning, collapse, and quantum revival stages. The linear PTC without Kerr nonlinearity provides the two-mode-squeezing reference, while Kerr nonlinearity introduces an anharmonic Fock-space potential that yields a turning point $n_{\mathrm{tp}}$, leading to collapse and revival behavior. (f) Mean biphoton number $\langle n \rangle$ versus evolution time, with marked points corresponding to the snapshots in b–e.

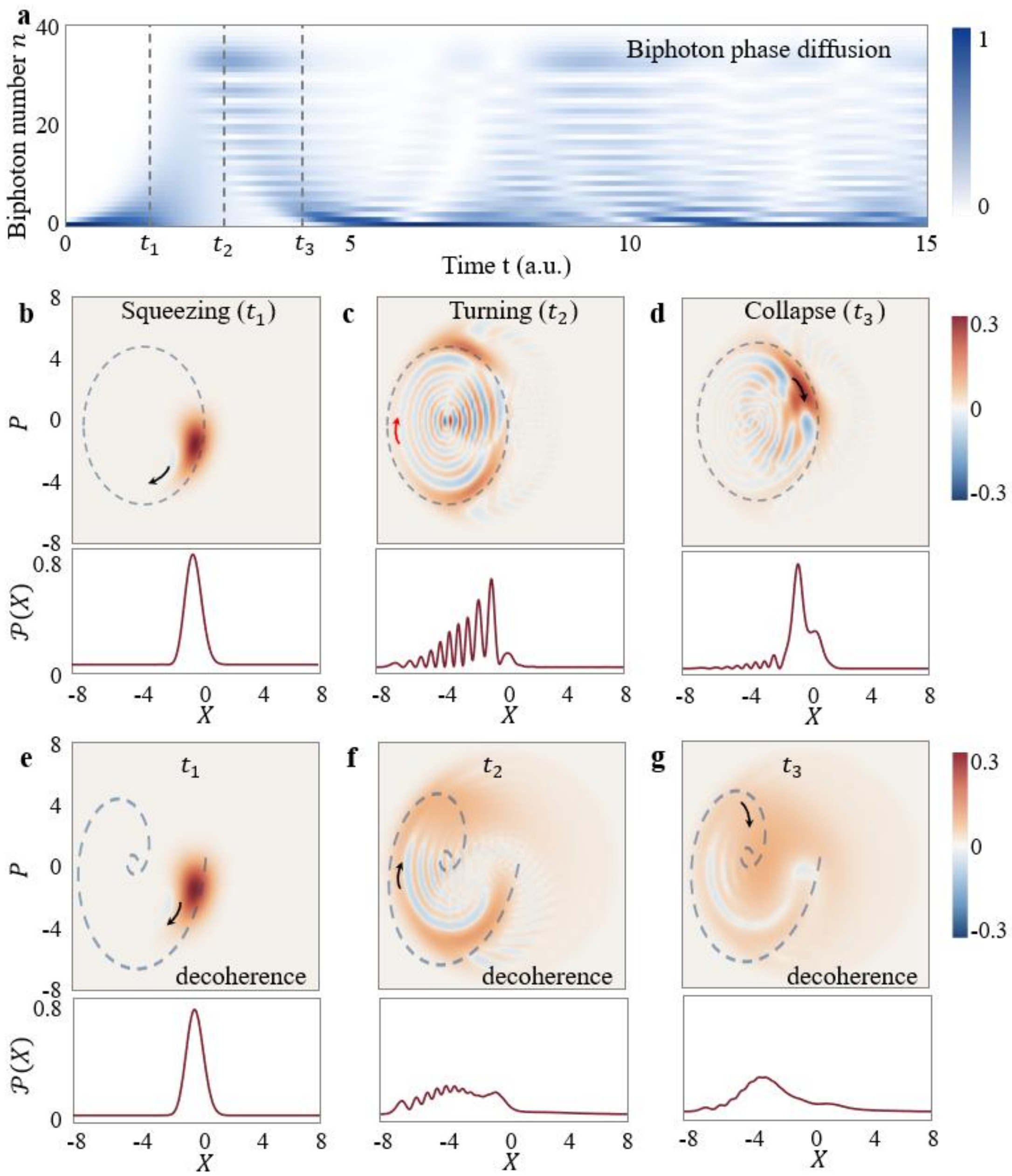

**Figure 3 | Biphoton phase-space representation of quantized *k*-gap solitons.** (a) Biphoton-number distribution $P(n,t)$ as a function of time. Dashed lines mark representative times $t_1$, $t_2$, and $t_3$, selected from the squeezing, turning, and collapse stages. At later times, Kerr-induced relative phase accumulation spreads the Fock-state components, leading to biphoton phase diffusion. (b–d) Ideal phase-space distributions in biphoton quadrature coordinates $(X,P)$, with the corresponding X-quadrature projections $\mathcal{P}(X)$ shown below. (e–g) Phase-space distributions and $X$-quadrature projections in the presence of photon loss and dephasing.

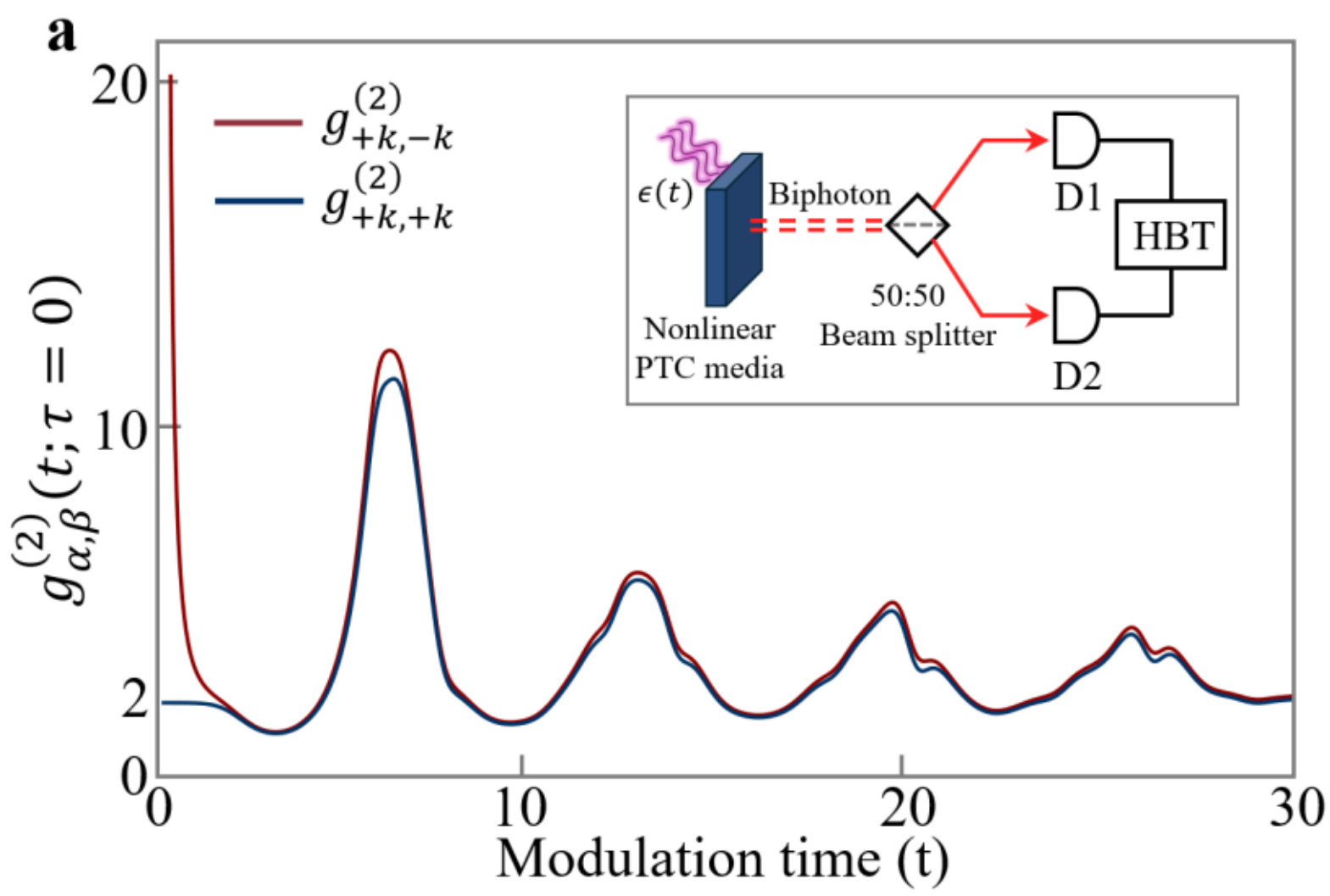

**Figure 4 | Hanbury Brown-Twiss measurements of quantized k-gap soliton dynamics.** (a) Zero-delay second-order correlations ($g^{(2)}$) as functions of modulation time $t$, showing Kerr-induced collapse and revival dynamics of the k-gap solitons. Inset: schematic HBT configuration, where the biphoton field is divided by a 50:50 beam splitter and measured by two single-photon detectors $D_1$ and $D_2$.

# Supplementary Information

# Quantization and Biphoton statistics of k-gap Solitons in Nonlinear Photonic Time Crystals

Liang Zhang, Chenhao Pan, Yiming Pan*

State Key Laboratory of Quantum Functional Materials, School of Physical Science and Technology and Center for Transformative Science, ShanghaiTech University, Shanghai 200031, China

The Supplementary Information provides detailed derivations, numerical definitions, and additional dynamical examples supporting the main text.

Supplementary Note 1 derives the classical coupled-mode theory.

Supplementary Note 2 justifies the biphoton description of quantized k-gap soliton dynamics.

Supplementary Note 3 presents the derivation of the nonlinear PTC Hamiltonian and the quantum Fock-ladder equation.

Supplementary Note 4 provides detailed Wigner quadrature phase-space distributions for both ideal and loss–dephasing dynamics.

Supplementary Note 5 presents coherent-state-initiated dynamics and compares it with the vacuum-initiated case discussed in the main text.

Supplementary Note 6 clarifies the algebraic structure of the biphoton Fock ladder and defines the effective quadrature operators used in the Wigner representation.

## 1. Classical coupled-mode theory

We derive the classical two-mode model from Maxwell's equations for a non-magnetic, source-free ($\rho = 0, \boldsymbol{J} = 0$) dielectric medium with an instantaneous Kerr nonlinearity. Applying the curl operator to Faraday's law and substituting Ampere's law yields the wave equation:

$$\frac{\partial^2 \boldsymbol{D}}{\partial t^2} = -\frac{1}{\mu_0} \nabla \times (\nabla \times \boldsymbol{E}). \tag{S1}$$

Assuming an optical Kerr effect for definiteness, the constitutive relation in a 1D isotropic medium is given by $\boldsymbol{D} = \epsilon(t, |\boldsymbol{E}|^2)\boldsymbol{E} = \epsilon_0[\epsilon_1(t) + \chi_3 |\boldsymbol{E}|^2]\boldsymbol{E}$ with $\chi_3$ being the Kerr coefficient. The linear dielectric constant is chosen to be spatially homogeneous but periodically modulated in time

$$\epsilon_1(t) = \epsilon_r(1 + \Delta\epsilon \cos \Omega t), \tag{S2}$$

where $\epsilon_r$ is the static relative permittivity, $\Delta\epsilon \ll 1$ is the dimensionless modulation depth (e.g., up to 0.3) and $\Omega = 2\pi/T$ the modulation frequency.

To proceed, we invert the constitutive relation to expand the electric field in terms of the displacement field. Dropping the vector notation for the 1D case, we obtain $\boldsymbol{E} = \frac{\boldsymbol{D}}{\epsilon_0 \epsilon_1} - \frac{\chi_3 \boldsymbol{D}^3}{\epsilon_0^2 \epsilon_1^4}$. Crucially, we formulate the nonlinear wave equation strictly in terms of $\boldsymbol{D}$ rather than $\boldsymbol{E}$. In a temporal crystal, the displacement field $\boldsymbol{D}$ guarantees rigorous continuity across temporal boundaries, a fundamental condition not necessarily satisfied by $\boldsymbol{E}$ . Substituting this expansion into Eq. (1) yields:

$$\frac{\partial^2 D}{\partial t^2} = \frac{1}{\mu_0} \frac{\partial^2}{\partial x^2} \left( \frac{D}{\epsilon_0 \epsilon_1} - \frac{\chi_3 D^3}{\epsilon_0^2 \epsilon_1^4} \right) \tag{S3}$$

Given the perturbative modulation depth ($\Delta\epsilon \ll 1$), we safely neglect its dynamic contribution to the higher-order Kerr term. Retaining the dominant self-phase modulation contribution, the wave equation simplifies to:

$$\frac{1}{c^2} \frac{\partial^2 D}{\partial t^2} = (1 - \Delta\epsilon \cos \Omega t) \frac{\partial^2 D}{\partial x^2} - \beta |D|^2 \frac{\partial^2 D}{\partial x^2} \tag{S4}$$

where $c = c_0/\sqrt{\epsilon_r} = c_0/n_0$ is the background speed of light and $\beta = 3\chi_3/\epsilon_0 \epsilon_r^3$ dictates the effective nonlinear response.

To capture the $k$-gap dynamics, we employ a coupled-mode framework based on the Floquet-Bloch theorem, constructing the field as a superposition of forward- and backward-propagating envelopes:

$$D(x,t) = A_f(x,t) e^{ikx - i\omega t} + \mathrm{A}_b(x,t) e^{ikx + i\omega t} + c.c., \tag{S5}$$

where $A_f$ and $A_b$ are the slowly varying amplitudes of forward- and backward-propagating modes, scattering with each other via modulation-induced time reflection. To maximize the scattering, the carrier frequency and momentum are phase-matched to the modulation such that $\omega = \Omega/2$ and $k = \Omega/2c$. Substituting the ansatz into Eq. (S4) and invoking the slowly-varying envelope approximation, we integrate out the fast-oscillating terms to obtain the governing nonlinear coupled-mode equations:

$$\begin{aligned} \frac{i}{c}\frac{\partial A_f}{\partial t} + i\frac{\partial A_f}{\partial x} + \kappa_L A_b + \gamma_0\left(\left|A_f\right|^2 + 2|A_b|^2\right)A_f = 0 \\ -\frac{i}{c}\frac{\partial A_b}{\partial t} + i\frac{\partial A_b}{\partial x} + \kappa_L A_f + \gamma_0\left(|A_b|^2 + 2\left|A_f\right|^2\right)A_b = 0 \end{aligned} \quad (S6)$$

where $\kappa_L = \Delta\epsilon\,\Omega/8c$ defines the temporal coupling coefficient and $\gamma_0 = \beta\Omega/4c$ represents the effective Kerr coefficient. Here $\kappa_L$ and $\gamma_0$ are per-length coefficients, as appropriate for the coupled-mode equations written with both $(1/c)\partial_t$ and $\partial_x$ .

To focus on the k-gap dynamics selected by temporal amplification, we restrict Eq. (S6) to a single phase-matched mode pair. Specifically, we consider spatially homogeneous envelopes at a fixed wavevector,

$$\frac{\partial A_f}{\partial x} = 0,\ \frac{\partial A_b}{\partial x} = 0. \quad (S7)$$

This approximation isolates one $(k, -k)$ mode pair and neglecting envelope propagation in space. For the intermediate projection, it is convenient to introduce the rescaled variable $\tau = ct_{\text{phys}}$, where $t_{\text{phys}}$ is the physical time. We further rewrite the forward and backward envelopes
as

$$A_f(\tau) \equiv A_k(\tau),\ A_b(\tau) \equiv -A_{-k}^*(\tau). \quad (S8)$$

This transformation converts the time-reflection coupling between forward and backward waves into a pair process between the $k$ and $-k$ modes. The minus sign is only a phase convention, chosen so that the temporal coupling appears with a positive coefficient.

After setting $\partial_x A_f = \partial_x A_b = 0$, Eq. (S6) becomes an ordinary two-mode equation in physical time. Therefore, to match the convention used in the main text, the per-length coefficients are converted into frequency-like coefficients by multiplying by $c$. We define

$$\delta(k) = c(|k| - k_0), \kappa = c\kappa_L = \frac{\Delta\epsilon\Omega}{8}, \gamma = -c\gamma_0 = -\frac{\beta\Omega}{4}. \quad (S9)$$

Here $\delta(k)$ is the residual detuning from the k-gap center, $\kappa$ is the temporal-modulation coupling used in the main text, and $\gamma$ is the signed effective Kerr coefficient. With these definitions, the reduced two-mode Kerr-PTC model becomes

$$i\frac{dA_k}{dt} = [\delta + \gamma(|A_k|^2 + 2|A_{-k}|^2)]A_k + \kappa A_{-k}^*$$

$$i\frac{dA_{-k}}{dt} = [\delta + \gamma(|A_{-k}|^2 + 2|A_k|^2)]A_{-k} + \kappa A_k^* \tag{S10}$$

This is the microscopic derivation of the reduced two-mode model used as Eq. (1) in the main text.

**2. Biphoton description of quantized k-gap soliton dynamics**

In the main text, we describe the quantized k-gap soliton dynamics using the selected opposite-momentum pair $(k_0, -k_0)$. Here we clarify why this biphoton description is appropriate. This reduction is based on the mode selection caused by linear k-gap amplification.

We introduce the wave-vector deviation $q = k - k_0$. So that $q = 0$ corresponds to the gap-center mode $k = k_0$, and the corresponding opposite-momentum pair is $(k_0, -k_0)$. inside the k-gap, the linear k-gap growth rate is $\Gamma(q) = \sqrt{\kappa^2 - \delta(q)^2}$, where $|\delta(q)| < \kappa$. Around the gap center, we use a linearized detuning, $\delta(q) = v_\delta q$. As shown in Fig. S1a. The maximum appears at $q = 0, \Gamma_{max} = \Gamma(0) = \kappa$. The gap-center mode has the largest linear k-gap amplification rate. Therefore, the phase-matched opposite-momentum pair $(k_0, -k_0)$ is preferentially selected by the k-gap.

The relative suppression of an off-center mode is quantified by

$$S(q,t) = \frac{I(q,t)}{I(0,t)} \simeq \exp^{-2(\Gamma_{max} - \Gamma(q))t} \ , \tag{S11}$$

Where $I(q,t) \propto e^{2\Gamma(q)t}$. Fig. S1b shows that $S(q,t)$ become progressively narrower with increasing time, meaning that modes away from $q = 0$ are increasingly suppressed during k-gap amplification.

We further monitor how much of the amplified field is carried by the selected pair. We quantify the selected-pair contribution by grouping the two opposite-momentum modes $+k_0 + q$ and $-k_0 - q$. We use the modal spectral weight

$$W_q(t) = |A_q|^2 + |B_q|^2, \tag{S12}$$

where $A_q$ and $B_q$ are the corresponding to the mode amplitudes. The selected pair fraction is defined as

$$R_b(t) = \frac{W_{q=0}(t)}{\sum_q W_q(t)}. \tag{S13}$$

The remaining off-center-mode fraction is

$$R_n(t) = 1 - R_b(t) = \frac{\sum_{q\neq 0} W_q(t)}{\sum_q W_q(t)}. \tag{S14}$$

Fig. S1c shows that $R_b(t)$ increases while $R_n(t)$ decreases for a weakly seeded gap-center pair with broadband noise. This confirms that k-gap amplification concentrates the amplified field into the selected $(k_0, -k_0)$ pair.

Therefore, the quantized k-gap soliton can be described by the biphoton dynamics of the selected opposite-momentum pair. After quantization, this pair forms the Fock ladder

$$|0,0\rangle,\ \ |1,1\rangle,\ \ |2,2\rangle, \dots,\ \ |n,n\rangle \dots.$$

which is the biphoton basis used in the main text.

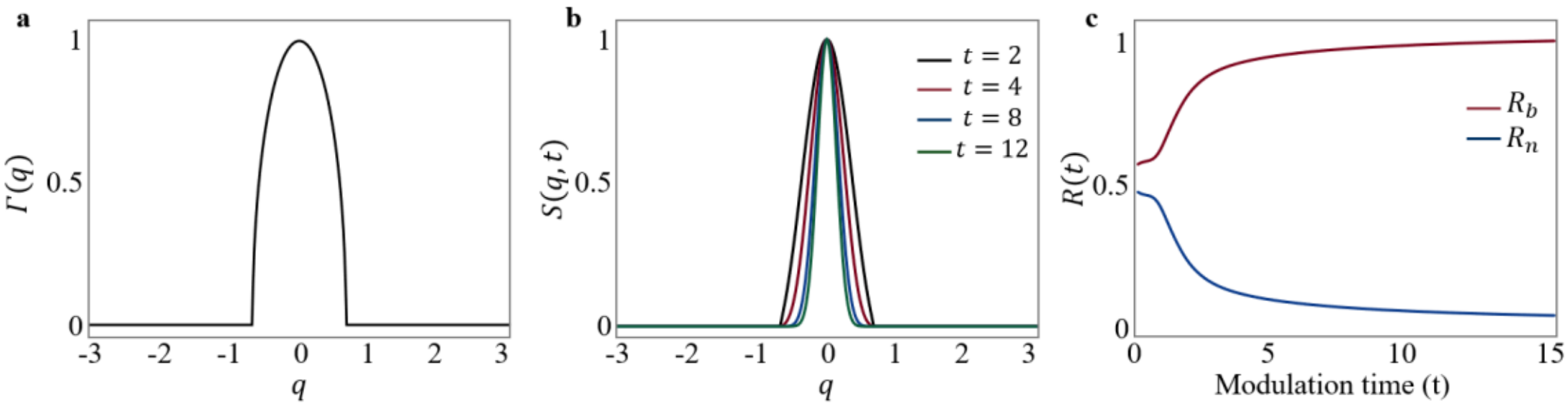

**Fig. S1 | Selection of the phase-matched k-gap pair.** (a) Linear growth rate $\Gamma(q)$ of k-gap modes, where $q = k - k_0$ measures the deviation from the gap center. (b) Relative suppression factor $S(q,t)$, showing that modes away from the gap center are increasingly suppressed during k-gap amplification. (c) Evolution of the phase-matched pair fraction $R_b$ and the remaining off-center-mode fraction $R_n$, supporting the two-mode reduction to the $(k_0, -k_0)$ pair.

**3. Derivation of the Hamiltonian and Fock-ladder equation**

We now construct a quantum model whose mean-field limit reproduces the reduced two-mode Kerr-PTC equations in Eq. (S10). The classical amplitudes $A_{\pm k}$ carry the dimension of field amplitude, whereas the bosonic creation and annihilation operators are dimensionless. We therefore introduce a real photon mode-amplitude scale $A_0$, and write

$$A_k \ \rightarrow A_0\, \hat{a}_k, A_k^* \ \rightarrow \ A_0 \hat{a}_k^\dagger,$$

$$A_{-k} \ \rightarrow A_0^*\, \hat{a}_{-k}, A_{-k}^* \ \rightarrow \ A_0^* \hat{a}_{-k}^\dagger, \tag{S15}$$

where $A_0$ is taken to be real, $A_0 = A_0^*$ is the quantized mode strength. Since the coupled-mode derivation is formulated in terms of the displacement-field amplitude, $A_0$ may be identified with the photon displacement-field scale of the selected mode. For a homogeneous nondispersive background with relative permittivity $\epsilon_r$ , the usual single-photon electric-field amplitude is

$$E_0 = \sqrt{\frac{\hbar\omega_0}{2\epsilon_0\epsilon_r V_{\text{eff}}}}, \tag{S16}$$

where $V_{\text{eff}}$ is the effective mode volume. Using $D = \epsilon_0 \epsilon_r E$, the corresponding displacement-field amplitude is

$$A_0 \equiv D_0 = \epsilon_0 \epsilon_r E_0 = \sqrt{\frac{\hbar\omega_0 \epsilon_0 \epsilon_r}{2V_{\text{eff}}}}. \qquad (S17)$$

With canonical commutation relations

$$[\hat{a}_j, \hat{a}_l^\dagger] = \delta_{jl}, \qquad \hat{n}_j = \hat{a}_j^\dagger \hat{a}_j. \qquad (S18)$$

The classical mean-field amplitudes are recovered as

$$A_j = A_0 \alpha_j, \qquad \alpha_j = \langle \hat{a}_j \rangle. \qquad (S19)$$

Thus, the classical intensity is related to the photon number by

$$| A_j |^2 = A_0^2 | \alpha_j |^2 \sim A_0^2 n_j. \qquad (S20)$$

This relation fixes the conversion between the classical nonlinear coefficient and the quantum Kerr coefficient. In the mean-field limit, the nonlinear term

$$\gamma(| A_j |^2 + 2 | A_l |^2) A_j$$

becomes

$$\gamma A_0^3 (| \alpha_j |^2 + 2 | \alpha_l |^2) \alpha_j.$$

However, the left-hand side of the classical equation also contains the same quantized mode strength,

$$i \frac{dA_j}{dt} = i A_0 \frac{d\alpha_j}{dt}. \qquad (S21)$$

After dividing the full equation by $A_0$, the nonlinear coefficient becomes

$$U = \gamma A_0^2 = \gamma \frac{\hbar\omega_0 \epsilon_0 \epsilon_r}{2V_{\text{eff}}}. \qquad (S22)$$

The temporal-modulation coefficient is not rescaled by $A_0$ under pair generation. We denote the corresponding quantum pair-coupling coefficient by $\tilde{\kappa}$, with

$$\tilde{\kappa} = \kappa. \qquad (S23)$$

With these definitions, the operator equations are required to reproduce Eq. (S10), and therefore take the form

$$i\frac{d\hat{a}_k}{dt} = \delta\hat{a}_k + \tilde{\kappa}\hat{a}_{-k}^{\dagger} + U(\hat{n}_k + 2\hat{n}_{-k})\hat{a}_k$$
$$i\frac{d\hat{a}_{-k}}{dt} = \delta\hat{a}_{-k} + \tilde{\kappa}\hat{a}_k^{\dagger} + U(\hat{n}_{-k} + 2\hat{n}_k)\hat{a}_{-k} \quad (S24)$$

The effective Hamiltonian is defined through the Heisenberg equation

$$i\hbar\frac{d\hat{a}_j}{dt} = [\hat{a}_j, \hat{H}_{\text{eff}}]. \quad (S25)$$

The detuning term is straightforward. To generate the linear term $\delta\hat{a}_k$ and $\delta\hat{a}_{-k}$, we choose

$$\hat{H}_\delta = \hbar\delta(\hat{n}_k + \hat{n}_{-k}). \quad (S26)$$

Indeed,

$$[\hat{a}_k, \hat{H}_\delta] = \hbar\delta[\hat{a}_k, \hat{n}_k] = \hbar\delta\hat{a}_k, \quad (S27)$$

and similarly

$$[\hat{a}_{-k}, \hat{H}_\delta] = \hbar\delta\hat{a}_{-k} \quad (S28)$$

Next, the temporal modulation term in Eq. (S10) contains $A_{-k}^*$ in the equation for $A_k$, and $A_k^*$ in the equation for $A_{-k}$. Therefore, in the quantum theory it should generate $\hat{a}_{-k}^{\dagger}$ and $\hat{a}_k^{\dagger}$. This is produced by the two-mode pair Hamiltonian

$$\hat{H}_{\tilde{\kappa}} = \hbar\tilde{\kappa}(\hat{a}_k^{\dagger}\,\hat{a}_{-k}^{\dagger} + \hat{a}_k\,\hat{a}_{-k}). \quad (S29)$$

Using the bosonic commutation relations,

$$[\hat{a}_k, \hat{a}_k^{\dagger}\,\hat{a}_{-k}^{\dagger}] = [\hat{a}_k, \hat{a}_k^{\dagger}\,]\hat{a}_{-k}^{\dagger} + \hat{a}_k^{\dagger}[\hat{a}_k, \hat{a}_{-k}^{\dagger}] = \hat{a}_{-k}^{\dagger}, \quad (S30)$$

and

$$[\hat{a}_k, \hat{a}_k\hat{a}_{-k}] = 0, \quad (S31)$$

We obtain

$$[\hat{a}_k, \hat{H}_{\tilde{\kappa}}] = \hbar\tilde{\kappa}\hat{a}_{-k}^{\dagger}. \quad (S32)$$

Thus $\hat{H}_{\tilde{\kappa}}$ generates the temporal pair-creation and pair-annihilation term. This is different from a beam-splitter interaction, which would have the form $\hat{a}_k^{\dagger}\hat{a}_{-k} + h.c.$. Here the PTC coupling creates or annihilates photons in opposite-momentum modes simultaneously.

We now construct the Kerr part. From Eq. (S24), the nonlinear contribution to the k-mode equation must be

$$U(\hat{n}_k + 2\hat{n}_{-k})\hat{a}_k.$$

This contains a self-phase modulation part $U\hat{n}_k\hat{a}_k$ and a cross-phase modulation part $2U\hat{n}_{-k}\hat{a}_k$. For the self-Kerr contribution, choose the normally ordered interaction

$$\hat{H}_{SPM,k} = \frac{\hbar U}{2}\hat{a}_k^{\dagger 2}\hat{a}_k^2. \quad (S33)$$

Using

$$[\hat{a}_k, \hat{a}_k^{\dagger 2}] = 2\hat{a}_k^\dagger, [\hat{a}_k, \hat{a}_k^2] = 0, \quad (S34)$$

we find

$$[\hat{a}_k, \hat{a}_k^{\dagger 2}\hat{a}_k^2] = 2\hat{a}_k^\dagger\hat{a}_k^2. \quad (S35)$$

Since

$$\hat{a}_k^\dagger\hat{a}_k^2 = \hat{a}_k^\dagger\hat{a}_k\hat{a}_k = \hat{n}_k\hat{a}_k, \quad (S36)$$

We obtain

$$[\hat{a}_k, \hat{H}_{SPM,k}] = \hbar U\hat{n}_k\hat{a}_k. \quad (S37)$$

Therefore $\hat{H}_{SPM,k}$ generates the self-Kerr term of the $k$-mode equation. Similarly,

$$\hat{H}_{\mathrm{SPM},-k} = \frac{\hbar U}{2}\hat{a}_{-k}^{\dagger 2}\hat{a}_{-k}^2 \quad (S38)$$

generates $U\hat{n}_{-k}\hat{a}_{-k}$ in the $-k$-mode equation.

For the cross-Kerr contribution, we need

$$2U\hat{n}_{-k}\hat{a}_k$$

In the k-mode equation and

$$2U\hat{n}_k\hat{a}_{-k}$$

in the $-k$-mode equation. This is generated by

$$\hat{H}_{XPM} = 2\hbar U\hat{n}_k\hat{n}_{-k}. \quad (S39)$$

Indeed,

$$[\hat{a}_k, \hat{H}_{XPM}] = 2\hbar U[\hat{a}_k, \hat{n}_k\hat{n}_{-k}]. \quad (S40)$$

Since $\hat{a}_k$ commutes with $\hat{n}_{-k}$,

$$[\hat{a}_k, \hat{n}_k\hat{n}_{-k}] = [\hat{a}_k, \hat{n}_k]\hat{n}_{-k} = \hat{a}_k\hat{n}_{-k}. \quad (S41)$$

Because $\hat{a}_k$ and $\hat{n}_{-k}$ act on different modes, they commute, so

$$\hat{a}_k\hat{n}_{-k} = \hat{n}_{-k}\hat{a}_k. \quad (S42)$$

Thus

$$\left[\hat{a}_k, \hat{H}_{XPM}\right] = 2\hbar U \hat{n}_{-k} \hat{a}_k. \tag{S43}$$

Similarly,

$$\left[\hat{a}_{-k}, \hat{H}_{XPM}\right] = 2\hbar U \hat{n}_k \hat{a}_{-k}. \tag{S44}$$

Combining the self- and cross-Kerr parts gives

$$\hat{H}_U = \frac{\hbar U}{2}\left(\hat{a}_k^{\dagger 2}\hat{a}_k^2 + \hat{a}_{-k}^{\dagger 2}\hat{a}_{-k}^2\right) + 2\hbar U \hat{n}_k \hat{n}_{-k}. \tag{S45}$$

Equivalently,

$$\hat{H}_U = \frac{\hbar U}{2}\left(\hat{a}_k^{\dagger 2}\hat{a}_k^2 + \hat{a}_{-k}^{\dagger 2}\hat{a}_{-k}^2 + 4\hat{n}_k \hat{n}_{-k}\right). \tag{S46}$$

Using

$$\hat{a}_j^{\dagger 2}\hat{a}_j^2 = \hat{n}_j\left(\hat{n}_j - 1\right) = \hat{n}_j^2 - \hat{n}_j, \tag{S47}$$

Eq. (S46) can also be written as

$$\hat{H}_U = \frac{\hbar U}{2}\left[\hat{n}_k^2 + \hat{n}_{-k}^2 + 4\hat{n}_k\, \hat{n}_{-k} - (\hat{n}_k + \hat{n}_{-k})\right]. \tag{S48}$$

This form makes explicit that the Kerr nonlinearity becomes a number-dependent energy shift in Fock space. Finally, the full effective Hamiltonian is

$$\hat{H}_{\text{eff}} = \hbar\delta(\hat{n}_k + \hat{n}_{-k}) + \hbar\tilde{\kappa}\left(\hat{a}_k^\dagger\, \hat{a}_{-k}^\dagger + \hat{a}_k\, \hat{a}_{-k}\right) + \frac{\hbar U}{2}\left[\hat{n}_k^2 + \hat{n}_{-k}^2 + 4\hat{n}_k\, \hat{n}_{-k} - (\hat{n}_k + \hat{n}_{-k})\right] \tag{S49}$$

This is the effective two-mode quantum Hamiltonian used as Eq. (3) in the main text. By construction, Eq. (S49) generates the operator equations (S24) through the Heisenberg equation. In the mean-field limit,

$$\hat{a}_j \to A_j, \hat{n}_j\hat{a}_j \to \left|A_j\right|^2 A_j, \tag{S50}$$

Eq. (S24) reduces back to the classical two-mode Kerr-PTC equations in Eq. (S10).

Because the pair term creates or annihilates photons simultaneously in the two modes, the photon-number difference

$$\hat{Q} = \hat{n}_k - \hat{n}_{-k} \tag{S51}$$

Is conserved,

$$\left[\hat{Q}, \hat{H}_{\text{eff}}\right] = 0. \tag{S52}$$

Therefore, if the system starts from the vacuum state $|0,0\rangle$, the dynamics remains in the $Q = 0$ sector,

$$|0,0\rangle,\ |1,1\rangle,\ |2,2\rangle, \dots, |n,n\rangle, \dots$$

We expand the state as

$$|\psi(t)\rangle = \sum_{n=0}^{\infty} c_n(t)|n,n\rangle\,, \tag{S53}$$

where $n$ denotes the photon-pair number.

We now evaluate the action of $\hat{H}_{\mathrm{eff}}$ on the basis state $|n,n\rangle$. The detuning term gives

$$\hbar\delta(\hat{n}_k + \hat{n}_{-k}) \mid n,n\rangle = 2\hbar\delta n \mid n,n\rangle. \tag{S54}$$

The Kerr term gives

$$\frac{\hbar U}{2}\left[\hat{n}_k^2 + \hat{n}_{-k}^2 + 4\hat{n}_k\,\hat{n}_{-k} - (\hat{n}_k + \hat{n}_{-k})\right]|n,n\rangle = \hbar U(3n^2 - n)|n,n\rangle. \tag{S55}$$

The pair-creation part gives

$$\hat{a}_k^{\dagger}\hat{a}_{-k}^{\dagger}|n,n\rangle = \left(\hat{a}_k^{\dagger} \otimes \hat{I}_{-k}\right)\left(\hat{I}_k \otimes \hat{a}_{-k}^{\dagger}\right)|n,n\rangle = (n+1)|n+1,n+1\rangle, \tag{S56}$$

because each creation operator contributes a factor $\sqrt{n+1}$. Similarly, the pair-annihilation part gives

$$\hat{a}_k\hat{a}_{-k}\,|n,n\rangle = \left(\hat{a}_k \otimes \hat{I}_{-k}\right)\left(\hat{I}_k \otimes \hat{a}_{-k}\right)|n,n\rangle = n|n-1,n-1\rangle. \tag{S57}$$

Therefore, this gives the pair-number ladder representation quoted as Eq. (4) in the main text.

$$\hat{H}_{\mathrm{eff}}|n,n\rangle = \hbar[2\delta n + U(3n^2 - n)]|n,n\rangle + \hbar\tilde{\kappa}(n+1)|n+1,n+1\rangle + \hbar\kappa n|n-1,n-1\rangle. \tag{S58}$$

Substituting

$$|\psi(t)\rangle = \sum_{n=0}^{\infty} c_n(t)|n,n\rangle\,, \tag{S59}$$

into the Schrödinger equation

$$i\hbar\frac{\partial}{\partial t}|\psi(t)\rangle = \hat{H}_{\mathrm{eff}}|\psi(t)\rangle, \tag{S60}$$

and collecting the coefficient of $|n,n\rangle$, we obtain

$$i\frac{dc_n}{dt} = [2\delta n + U(3n^2 - n)]c_n + \tilde{\kappa} n c_{n-1} + \kappa(n+1)c_{n+1} \tag{S61}$$

Equation (S61) is the full derivation of the pair-number amplitude equation used as Eq. (5) in the main text. Here $c_{-1} = 0$, and in numerical calculations with a finite cutoff $N_{max}$. This equation is the effective one-dimensional Fock-ladder equation for the quantized photon-pair dynamics. Throughout this derivation, the detuning term $\delta$ is retained for generality. In the main text, we focus on the k-gap center by setting $\delta = 0$ when appropriate.

**4. Wigner quadrature phase-space distributions under ideal and dissipative dynamics**

This section provides additional Wigner quadrature phase-space snapshots complementing the discussion in the main text. We use them to illustrate how the quantized k-gap soliton evolves in the effective biphoton-ladder phase space, and how its nonclassical interference is preserved or suppressed under ideal and dissipative dynamics.

Figure S2 shows the ideal evolution. At $t = 0$, the vacuum state is localized near the origin of the $(X, P)$ phase space and has a positive enveloped Wigner distribution. At early times, such as $t = 1$, the distribution becomes elongated and tilted, corresponding to the squeezing stage of the biphoton dynamics. In this regime, the biphoton Wigner function remain largely positive and close to a squeezed-state profile, consistent with the early-time behavior in Fig. 2b of the main text. At later times, the state becomes strongly non-Gaussian. From $t = 2$ to $t = 3$, the Wigner distribution moves away from the origin and develops oscillatory positive and negative regions, reflecting interference fringes between different biphoton Fock states. This behavior corresponds to the turning-stage behavior discussed in the main text, where Wigner negativity provides a direct indicator of nonclassicality. From $t = 3$ to $t = 8$, the distribution continues to deform and rotate in phase space, visualizing the squeezing–turning–collapse–revival cycle of the biphoton Fock-ladder. The quantized soliton state first stretches outward, reaches a turning region with pronounced interference fringes, collapses back toward lower pair-number states, and later revives. The recurrent nonclassical structure is quantified by the Wigner negativity in Fig. S2k, which shows repeated growth and suppression during the cycle.

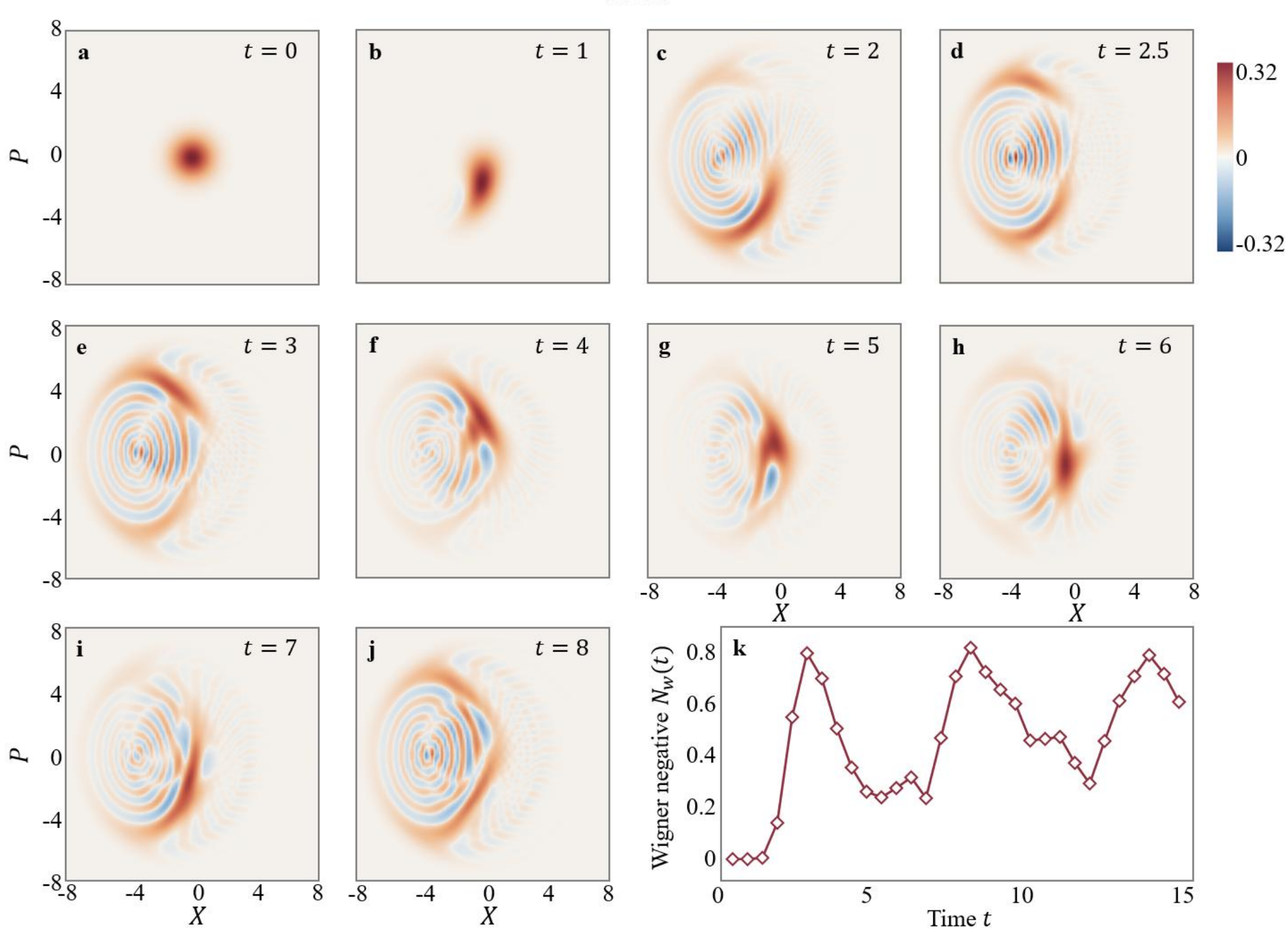

**Fig. S2 | Ideal Wigner quadrature phase-space dynamics.** (a–j) Wigner distributions $W(X,P)$ at representative times from $t=0$ to $t=8$ under ideal evolution. (k) Wigner negativity $N_w(t)$ showing the emergence and recurrence of nonclassical phase-space features.

Figure S3 shows the Wigner-quadrature evolution in the presence of photon loss and dephasing within the biphoton Fock ladder. We evolve the density matrix according to a Lindblad master equation,

$$\frac{d\rho}{dt} = -i[\hat{\mathcal{H}},\rho] + \Gamma_\phi \mathcal{D}[\hat{N}]\rho + \Gamma_{loss}\mathcal{D}[\hat{b}]\rho, \tag{S62}$$

where $\mathcal{D}[\hat{O}]\rho = \hat{O}\rho\hat{O}^\dagger - \frac{1}{2}\{\hat{O}^\dagger\hat{O},\rho\}$. $\hat{\mathcal{H}} = \frac{\hat{H}_{\text{eff}}}{\hbar}$, $\hat{N}$ is the photon-pair number operator, and $\hat{b}$ is the effective lowering operator. The rate $\Gamma_\phi$ describes dephasing between different pair-number states, while $\Gamma_{loss}$ describes loss. In the $|n\rangle$ basis, the combined dephasing and loss contribution is

$$\frac{d\rho_{mn}}{dt} = -\frac{\Gamma_\phi}{2}(m-n)^2\rho_{mn} + \Gamma_{loss}\left[\sqrt{(m+1)(n+1)}\rho_{m+1,n+1} - \frac{m+n}{2}\rho_{mn}\right]. \tag{S63}$$

The first term suppresses off-diagonal coherence between different pair-number states, while the second transfers population from higher to lower photon-pair states and damps the associated coherences. Thus, loss and dephasing together smooth the Wigner distribution by weakening both the high-n components and the phase coherence between them.

The resulting dissipative evolution is shown in Fig. S3. At early times, such as $t = 1$, the dissipative evolution remains close to the ideal case: the Wigner distribution stays near the origin and develops an elongated squeezed-state-like profile. This indicates that the initial squeezing stage is not immediately destroyed by the chosen loss and dephasing rates. The deviation becomes evident when fine interference structures emerge. At $t = 2$ and $t = 2.5$, the distribution still undergoes a broad phase-space deformation, but the positive–negative oscillations are much weaker than in the ideal case. By $t = 3$ and later, the interference fringes are strongly suppressed, and the Wigner distribution becomes smoother and mostly positive. This reflects loss of off-diagonal coherence between different biphoton-number states, which is responsible for the fringes and negative regions in phase space. Unlike the ideal case, the dissipative dynamics no longer shows a near-rotational return. Instead, the distribution relaxes around the center of a dissipatively modified trajectory. Consistently, the Wigner negativity in Fig. S3k shows only a small early-time peak and then decays close to zero.

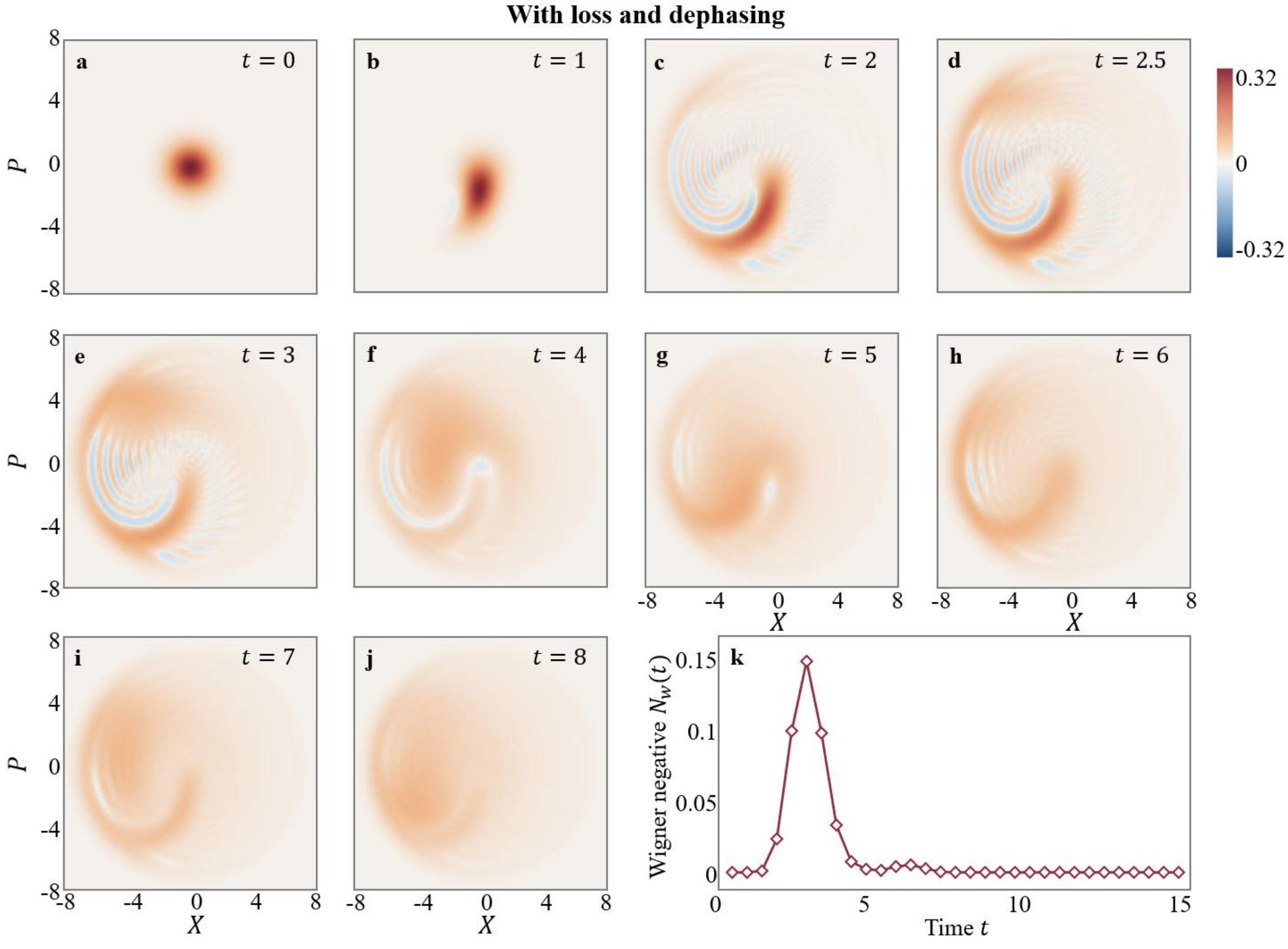

**Fig. S3 | Wigner quadrature phase-space dynamics with loss and dephasing.** (a–j) Wigner distributions $W(X, P)$ at representative times from $t = 0$ to $t = 8$, calculated under dissipative evolution with photon loss and dephasing. (k) Wigner negativity $N_w(t)$, showing that dissipation suppresses nonclassical phase-space interference after the early squeezing stage.

**5. Coherent-state-initiated dynamics**

In the main text, we focus on vacuum-initiated dynamics, where the system remains in the $Q = 0$ biphoton ladder, with $Q = n_k - n_{-k}$. Here we consider a complementary case initialized from coherent states in the two opposite-momentum modes,

$$|\psi(0)\rangle = |\alpha\rangle_k \otimes |\alpha\rangle_{-k}. \qquad (S64)$$

For the symmetric case studied here, the state can be written as

$$|\psi(0)\rangle = e^{-|\alpha|^2} \sum_{n_k, n_{-k}=0}^{\infty} \frac{\alpha^{n_k+n_{-k}}}{\sqrt{n_k!\, n_{-k}!}} \mid n_k, n_{-k}\rangle. \qquad (S65)$$

Unlike the vacuum state, this coherent initial state is not confined to a single Fock component or a single $Q = 0$ sector. Instead, it occupies a finite distribution in the two-dimensional photon-number space $(n_k, n_{-k})$. However, this two-dimensional problem can still be reorganized into a set of independent one-dimensional Fock ladders. Since the Hamiltonian conserves the photon-number difference $Q = n_k - n_{-k}$, the Hilbert space decomposes into independent $Q$-resolved sectors. Accordingly, the coherent initial state can be written as a coherent superposition of $Q$-resolved ladder components,

$$|\psi(0)\rangle = \sum_{Q=-\infty}^{\infty} |\psi_Q(0)\rangle, \qquad (S66)$$

where

$$|\psi_Q(0)\rangle = e^{-|\alpha|^2} \sum_{n=0}^{\infty} \frac{\alpha^{2n+|Q|}}{\sqrt{(n+Q_+)!\,(n+Q_-)!}} \mid n+Q_+, n+Q_-\rangle. \qquad (S67)$$

Here, $Q_+ = \max(Q, 0)\,, Q_- = \max(-Q, 0)$. Thus, the coherent-state dynamics can be viewed as parallel evolution on a series of independent $Q$-dependent Fock ladders. The $Q = 0$ ladder discussed in the main text is simply the special biphoton state $\mid n, n\rangle$, whereas the coherent initial state additionally populates $Q \neq 0$ ladders such as $|n+Q, n\rangle (Q > 0)$, and $|n, n+\mid Q \mid \rangle (Q < 0)$. Each $Q$ ladder evolves independently under its corresponding block Hamiltonian, while the full coherent-state dynamics is obtained by summing the contributions from all occupied $Q$ sectors.

Figure S3 shows the resulting joint photon-number distribution $P(n_k, n_{-k})$. Initially, the distribution is localized around the mean photon number set by $\alpha$ [Fig. S4a]. During the early evolution, for example at $t = 1$ [Fig. S4b,], it is amplified toward larger photon numbers and spreads mainly along the correlated diagonal direction $n_k \simeq n_{-k}$, corresponding to the squeezing stage of the seeded dynamics. The growth is then arrested by Kerr nonlinearity. Near the turning stage ($t = 2$) [Fig. S4c], the distribution bends back toward lower photon-number states, leading to the collapse stage at $t = 3$ [Fig. S4d]. After the collapse, the squeezing becomes effective again in the lower-photon-number region, and the distribution expands again [Figs. S4e–g]. The mean biphoton number $\langle n\rangle$ in Fig. S3h confirms this squeezing-turning-collapse-revival sequence. However, this cycling is damping due to the intrinsic phase diffusion of photon numbers. Thus, the collapse–revival dynamics of the quantized k-gap soliton is not restricted to vacuum initiation, but also persists under coherent-state seeding.

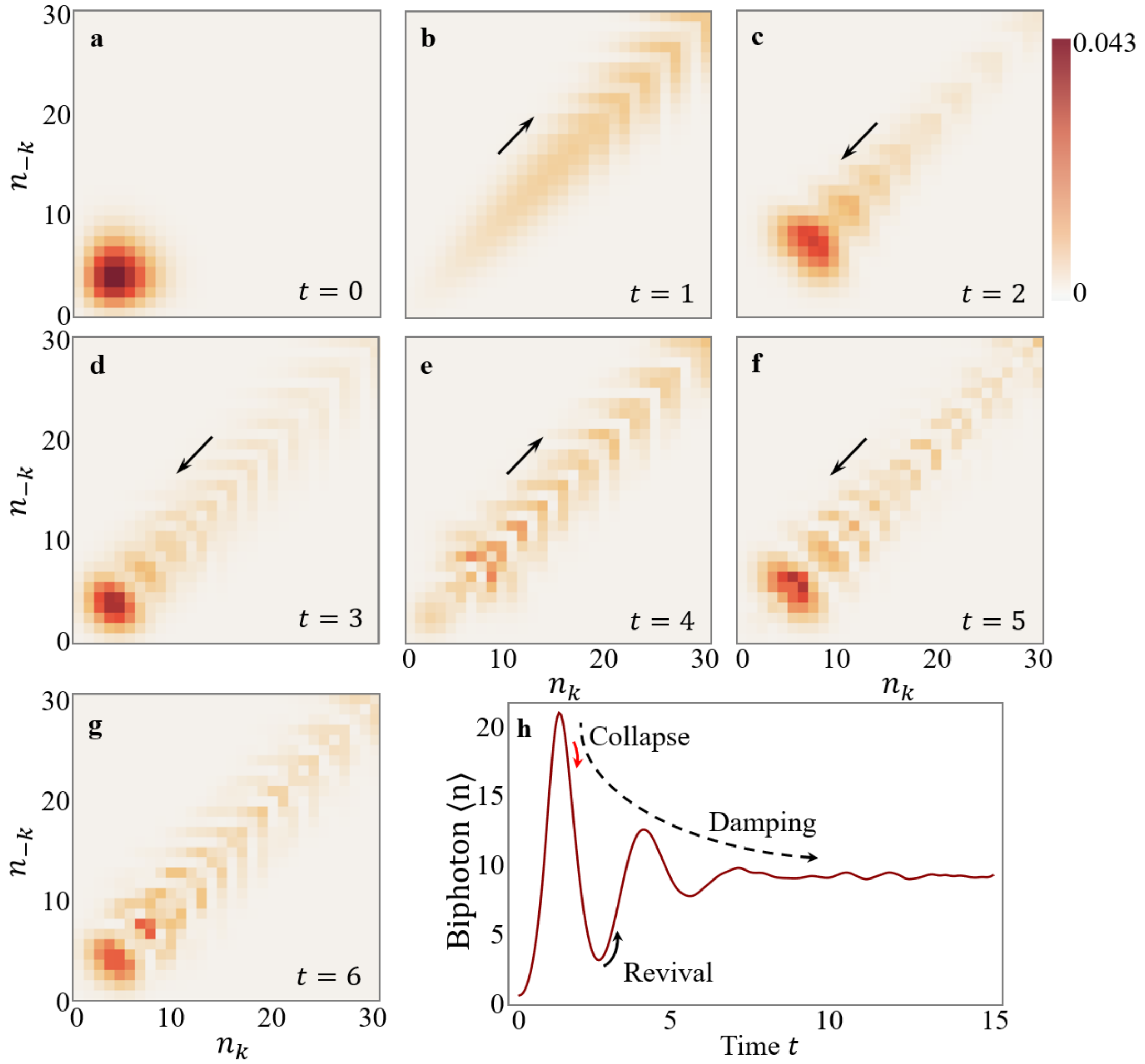

**Fig. S4 | Biphoton statistics in nonlinear photonic time crystals from a coherent-state initial condition.** (a–g) Joint photon-number distributions $P(n_k, n_{-k})$ at representative times from $t = 0$ to $t = 6$, initialized from a coherent state. (h) Mean biphoton number ⟨n⟩ as a function of time, showing oscillatory collapse–revival behavior of the quantized k-gap solitons.

**6. Algebraic structure and effective Wigner representation of the biphoton Fock ladder**

In the main text, the quantized $k$-gap soliton dynamics is represented on the biphoton Fock ladder. Here we clarify the algebraic structure of this ladder and the meaning of the effective quadrature variables $(X, P)$ used in the Wigner representation. The key point is that the physical pair process is a composite two-photon process, whereas the Wigner function is constructed using an canonical representation for visualization. Therefore, the "biphoton" used here should not be interpreted as a new elementary bosonic particle.

The two underlying optical modes are ordinary bosonic modes satisfying

$$[\hat{a}_j, \hat{a}_l^\dagger] = \delta_{jl}, \qquad j, l = \pm k. \tag{S68}$$

Temporal modulation in the photonic time crystal creates or annihilates photons simultaneously in the two opposite-momentum modes. We therefore define the composite pair operators

$$\hat{B}^{\dagger} = \hat{a}_k^{\dagger}\hat{a}_{-k}^{\dagger}, \qquad \hat{B} = \hat{a}_k\hat{a}_{-k}. \tag{S69}$$

These operators conserve the photon-number difference

$$\hat{Q} = \hat{n}_k - \hat{n}_{-k}, \tag{S70}$$

because

$$[\hat{Q}, \hat{B}^{\dagger}] = 0, \qquad [\hat{Q}, \hat{B}] = 0. \tag{S71}$$

For a vacuum initial state, the dynamics remains in the $Q = 0$ sector. We denote the correlated pair-number basis as

$$| n\rangle_B \equiv | n, n\rangle, \tag{S72}$$

where $n$ is the photon-pair number. In this work, the word "biphoton" refers to this correlated two-mode state $| n, n\rangle$, not to an elementary canonical boson.

The action of the physical pair operators on this ladder is

$$\hat{B}^{\dagger} | n\rangle_B = \hat{a}_k^{\dagger}\hat{a}_{-k}^{\dagger} | n, n\rangle = (n+1) | n+1\rangle_B, \tag{S73}$$

and

$$\hat{B} | n\rangle_B = \hat{a}_k\hat{a}_{-k} | n, n\rangle = n | n-1\rangle_B. \tag{S74}$$

Thus, the physical pair process has matrix elements $n+1$ and $n$. These are different from the $\sqrt{n+1}$ and $\sqrt{n}$ matrix elements of a canonical single-boson ladder. Equivalently,

$$[\hat{B}, \hat{B}^{\dagger}] = [\hat{a}_k\hat{a}_{-k}, \hat{a}_k^{\dagger}\hat{a}_{-k}^{\dagger}] = \hat{n}_k + \hat{n}_{-k} + 1. \tag{S75}$$

Within the $Q = 0$ sector, this becomes

$$[\hat{B}, \hat{B}^{\dagger}] | n\rangle_B = (2n+1) | n\rangle_B. \tag{S76}$$

Therefore, $\hat{B}, \hat{B}^{\dagger}$ do not obey the canonical bosonic commutation relation. This non-canonical composite-pair algebra is the origin of the $n$-dependent hopping amplitudes in the biphoton

Fock-ladder equation.

The Kerr interaction is diagonal in the same pair-number basis. In the $Q = 0$ sector, it acts as

$$\hat{H}_{\mathrm{U}} \mid n\rangle_B = \hbar U(3n^2 - n) \mid n\rangle_B. \tag{S77}$$

Combining the diagonal Kerr term with the composite pair creation and annihilation terms gives the effective one-dimensional ladder equation

$$i\frac{dc_n}{dt} = U(3n^2 - n)c_n + \tilde{\kappa}nc_{n-1} + \kappa(n+1)c_{n+1}. \tag{S78}$$

Equation (S74) shows explicitly that the quantized $k$-gap dynamics is not the dynamics of an elementary biphoton boson. Instead, it is the dynamics of a composite photon-pair ladder inherited from two ordinary bosonic modes.

We now define the effective quadrature variables used in the Wigner representation. To visualize the state on a phase space, we map the discrete ladder $\{\mid n\rangle_B\}$ onto the number-state basis of a one-dimensional harmonic oscillator. Specifically, we introduce a canonical ladder operator $\hat{b}$ by its action on the biphoton ladder basis,

$$\hat{b} \mid n\rangle_B = \sqrt{n} \mid n-1\rangle_B, \qquad \hat{b}^\dagger \mid n\rangle_B = \sqrt{n+1} \mid n+1\rangle_B. \tag{S79}$$

This operator satisfies

$$[\hat{b}, \hat{b}^\dagger] = 1.$$

The effective quadrature operators used in the Wigner representation are then defined as

$$\hat{X} = \frac{\hat{b} + \hat{b}^\dagger}{\sqrt{2}}, \qquad \hat{P} = \frac{\hat{b} - \hat{b}^\dagger}{i\sqrt{2}} \tag{S80}$$

so that

$$[\hat{X}, \hat{P}] = i. \tag{S81}$$

Equivalently, their matrix elements in the biphoton ladder basis are

$${}_B\langle m \mid \hat{X} \mid n\rangle_B = \frac{1}{\sqrt{2}}\left[\sqrt{n+1}\,\delta_{m,n+1} + \sqrt{n}\,\delta_{m,n-1}\right], \tag{S82}$$

and

$${}_B\langle m \mid \hat{P} \mid n\rangle_B = \frac{1}{i\sqrt{2}}\left[\sqrt{n+1}\,\delta_{m,n+1} - \sqrt{n}\,\delta_{m,n-1}\right]. \tag{S83}$$

The continuous variables $X$ and $P$ in Fig. 3 are the eigenvalue coordinates associated with these effective quadrature operators. They should not be interpreted as the real-space coordinate and physical momentum of the optical field. They are also not the quadratures of the composite pair operators $\hat{K}_{\pm}$. Instead, they are the canonical quadratures of the oscillator representation of the biphoton-number ladder.

The distinction between $\hat{b}$ and the physical pair operators is essential. On the $Q = 0$ ladder, the two are related by

$$\hat{B}^{\dagger} = \hat{b}^{\dagger}\sqrt{\hat{N}+1}, \hat{B} = \sqrt{\hat{N}+1}\,\hat{b}, \tag{S84}$$

where

$$\hat{N} \mid n\rangle_B = n \mid n\rangle_B.$$

Equation (S79) makes clear that $\hat{B}^{\dagger} \neq \hat{b}^{\dagger}$ and $\hat{B} \neq \hat{b}$. The physical dynamics is generated by $\hat{B}^{\dagger}$ and $\hat{B}$, whose matrix elements are $n+1$ and $n$, while the Wigner representation uses $\hat{b}^{\dagger}$ and $\hat{b}$ only to assign a canonical quadrature phase space to the pair-number ladder.

With this quadrature representation, the density matrix in the biphoton ladder can be written as

$$\hat{\rho}_B(t) = \sum_{n,m} \rho_{nm}(t) \mid n\rangle_{B\ B}\langle m \mid, \tag{S85}$$

where,

$$\rho_{nm}(t) = c_n(t)c_m^*(t). \tag{S86}$$

The effective Wigner function is defined as

$$W_B(X, P; t) = \sum_{n,m} \rho_{nm}(t)\, W_{nm}(X, P), \tag{S87}$$

where $W_{nm}(X, P)$ is the standard harmonic-oscillator Wigner representation of the operator $|n\rangle_{B\ B}\langle m|$ under the mapping in Eq. (S75).

The $X$-quadrature marginal distribution shown below the Wigner functions is

$$\mathcal{P}(X; t) = \int W_B(X, P; t)\, dP. \tag{S88}$$

The diagonal density-matrix elements $\rho_{nn} = \mid c_n \mid^2$ give the biphoton-number distribution $P(n, t)$. By contrast, the off-diagonal elements $\rho_{nm}$ with $n \neq m$ contain the relative phases between different biphoton-Fock states and generate the interference fringes and negative regions in $W_B(X, P; t)$. Thus, the Wigner function is used here as a diagnostic representation of coherence and interference in the biphoton Fock ladder, while the underlying dynamics remains governed by the non-canonical composite-pair algebra of the physical pair operators $\hat{B}$ and $\hat{B}^{\dagger}$.